\documentclass[aps,prb,twocolumn,superscriptaddress,floatfix]{revtex4}%
\usepackage{graphicx}
\usepackage{amsmath}
\usepackage{amsfonts}
\usepackage{amssymb}%
\def\nbZ{{\mathchoice {\hbox{$\sf\textstyle Z\kern-0.4em Z$}}
{\hbox{$\sf\textstyle Z\kern-0.4em Z$}} {\hbox{$\sf\scriptstyle
Z\kern-0.3em Z$}}  {\hbox{$\sf\scriptscriptstyle Z\kern-0.2em Z$}}}}

\newcommand{\rbf}{\ensuremath{\mathbf{r}}}

\newcommand{\be}{\begin{equation}}
\newcommand{\ee}{\end{equation}}
\begin{document}
\title{Non-Abelian Chern-Simons models with discrete gauge groups on a lattice }
\author{B. Dou\c{c}ot}
\affiliation{Laboratoire de Physique Th\'{e}orique et Hautes \'Energies, CNRS UMR 7589,
Universit\'{e}s Paris 6 et 7, 4, place Jussieu, 75252 Paris Cedex 05 France}
\author{L.B. Ioffe}
\affiliation{Center for Materials Theory, Department of Physics and Astronomy, Rutgers
University 136 Frelinghuysen Rd, Piscataway NJ 08854 USA}

\begin{abstract}
We construct the local Hamiltonian description of the Chern-Simons theory with
{\ discrete non-Abelian }gauge group on a lattice. We show that the theory is
fully determined by the phase factors associated with gauge transformations
and classify all possible non-equivalent phase factors. We also construct the
gauge invariant electric field operators that move fluxons around and
create/anihilate them. We compute the resulting braiding properties of the
fluxons. We apply our general results to the simplest class of non-Abelian
groups, dihedral groups $D_{n}$.

\end{abstract}
\maketitle

\section{Introduction.}

Realization of a quantum computer would be very important for the solution of
hard problems in classical mathematics \cite{Shor1994} and even more important
for the solution of many quantum problems in field theory, quantum chemistry,
material physics, etc.\cite{Feynman1996,Ekert1996,Steane1998} However, despite
a large effort of many groups the realization remains far away because of the
conflicting requirements imposed by scalability and decoupling from the
environment. This dichotomy can be naturally resolved by the error free
computation in the topologically protected space \cite{Dennis2002,Childs2002}
if the physical systems realizing such spaces are identified and implemented.
Though difficult, this seems more realistic than attempts to decouple simple
physical systems implementing individual bits from the environment. Thus, the
challenge of the error free quantum computation~ resulted in a surge of
interest to physical systems and mathematical models that were considered very
exotic before.

The main idea of the whole approach, due to Kitaev \cite{Kitaev1997}, is that
the conditions of long decoherence rate and scalability can be in principle
satisfied if elementary bits are represented by anyons, the particles that
indergo non-trivial transformations when moved adiabatically around each other
(braided)~\cite{Kitaev1997,Mochon2003,Mochon2004} and that can not be
distinguished by local operators. One of the most famous examples of such
excitations is provided by the fractional Quantum Hall
Effect~\cite{Halperin84,Arovas84}. The difficult part is, of course, to
identify a realistic physical system that has such excitations and allows
their manipulations. To make it more manageable, this problem should be
separated into different layers. The bottom layer is the physical system
itself, the second is the theoretical model that identifies the low energy
processes, the third is the mathematical model that starts with the most
relevant low energy degrees of freedom and gives the properties of anyons
while the fourth deals with construction of the set of rules on how to move
the anyons in order to achieve a set of universal quantum gates (further lies
the layer of algorithms and so on). Ideally, the study of each layer should
provide the one below it a few simplest alternatives and the one above the
resulting properties of the remaining low energy degrees of freedom.

In this paper we focus on the third layer: we discuss a particular set of
mathematical models that provides anyon excitations, namely the Chern Simons
gauge theories with the discrete gauge groups. Generally, an appealing
realization of the anyons is provided by the fluxes in non-Abelian gauge
theories. \cite{Mochon2003,Mochon2004}. The idea is to encode individual bits
in the value of fluxes belonging to the same conjugacy class of the gauge
group: such fluxes can not be distinguished locally because they are
transformed one into another by a global gauge transformation and would be
even completely indistiguishable in the absence of other fluxes in the system.
Alternatively, one can protect anyons from the adverse effect of the local
operators by spreading the fluxes over a large region of space. In this case
one does not need to employ a non-Abelian group: the individual bits can be
even encoded by the presence/absence of flux in a given area, for instance a
large hole in the lattice. Such models
\cite{Ioffe2002a,Ioffe2002b,Doucot2003,Doucot2004a,Doucot2004b} are much
easier to implement in solid state devices but they do not provide a large set
of manipulation that would be sufficient for universal quantum computation.
Thus, these models provide a perfect quantum memory but not a quantum
processor. On the other hand, the difficulty with the flux representation is
that universal computation can be achieved only by large non-Abelian groups
(such as $A_{5}$) that are rather difficult to implement, or if one adds
charge motion to the allowed set of manipulations. Because the charges form a
non-trivial representation of the local gauge group, it is difficult to
protect their coherence in the physical system which makes the latter
alternative also difficult to realize in physical systems. The last
alternative is to realize a Chern-Simons model where on the top of the
conjugacy transformations characteristic of non-Abelian theories, the fluxes
acquire non-trivial phase factors when moved around each other. We explore
this possibility in this paper.

Chern-Simons theories with finite non-Abelian gauge groups have been
extensively studied for a continuous $2+1$ dimensional space-time. Unlike
continous gauge-group, discrete groups on a continous space allow non-trivial
fluxes only around cycles which cannot be contracted to a single point. So in
a path-integral approach, and by contrast to the case of continuous groups,
the integration domain is restricted to gauge-field configurations for which
the local flux density vanishes everywhere. Such path integrals were
introduced, analyzed, and classified in the original paper by Dijkgraaf and
Witten~\cite{Dijkgraaf1990}. They showed that for a given finite gauge group
$G$, all possible Chern-Simons actions in $2+1$ dimensions are in one to one
correspondence with elements in the third cohomology group $H^{3}(G,U(1))$ of
$G$ with coefficients in $U(1)$. They also provided a description in terms of
a $2+1$ lattice gauge theory, where space-time is tiled by tetrahedra whose
internal states are described by just three independent group elements
$(g,h,k)$ because fluxes through all triangular plaquettes vanish. Elements of
$H^{3}(G,U(1))$ are then identified with functions \mbox{$\alpha(h,k,l)$} that
play the role of the elementary action for a tetrahedron. This description
turns out to be rather cumbersome because \mbox{$\alpha(h,k,l)$} does not have
specific symmetry properties, so that the definition of the total action
requires to choose an ordering for all the lattice sites, which cannot be done
in a natural way. As a result, it seems very difficult to take the limit of a
continuous time that is necessary for our purposes because physical
implementations always require an explicit Hamiltonian form.

In principle, the knowledge of an elementary action $\alpha$ in $H^{3}%
(G,U(1))$ allows to derive all braiding properties of anyonic excitations
(i.e. local charges and fluxes). This has been done directly from the original
path-integral formulation~\cite{Freed93,Freed94}, using as an intermediate
step the representation theory of the so-called quasi-quantum double
associated to the group $G$~\cite{DPR90}. This mathematical struture is
completely defined by the group $G$ and a choice of any element $\alpha$ in
$H^{3}(G,U(1))$. Using this later formalism, which bypasses the need to
construct microscopic Hamiltonians, many detailed descriptions of anyonic
properties for various finite groups, such as direct products of cyclic
groups, or dihedral groups, have been given by de Wild
Propitius~\cite{Propitius95}. Unfortunately, these general results do not
provide a recipe how to contruct a microscopic Hamiltonian with prescribed
braiding properties.

To summarize, in spite of a rather large body of theoretical knowledge, we
still do not know which Chern-Simons models with a finite gauge group can
\emph{in principle} be realized in a physical system with a local Hamiltonian,
that is which one can appear as low energy theory. To answer this question is
the purpose of the present paper.

The main ingredients of our construction are the following. We shall be mostly
interested in a Hilbert space spanned by \emph{dilute} classical
configurations of fluxes because it is only these configurations that are
relevant for quantum computation that involves flux braiding and fusion.
Furthermore, we expect that a finite spacial separation between fluxons may
facilitate various manipulations and measurements. Notice in this respect that
in theories with discrete group symmetry in a continuous
space\cite{Dijkgraaf1990,Freed93,Freed94}, non-trivial fluxes appear only
thanks to a non-trivial topology of the ambiant space~\cite{steenrod1951} and
thus are restricted to large well separated holes in a flat 2D space. The
second feature of our construction is that gauge generators are modified by
phase-factors which depend on the actual flux configuration in a local way.
This is a natural generalization of the procedure we have used recently for
the construction of Chern-Simons models with
${\mathchoice {\hbox{$\sf\textstyle Z\kern-0.4em Z$}}{\hbox{$\sf\textstyle Z\kern-0.4em Z$}}{\hbox{$\sf\scriptstyle
Z\kern-0.3em Z$}}{\hbox{$\sf\scriptscriptstyle Z\kern-0.2em Z$}}}_{N}$
symmetry group~\cite{Doucot2005,Doucot2005b}. Note that in the most general
situations, the representation of the local gauge group in a discrete
Chern-Simons theory becomes \emph{projective}~\ \cite{Dijkgraaf1990,
DPR90,Freed93,Freed94,Propitius95}. This would be inappropriate for a robust
implementation because projective representations lead to degenerate
multiplets that are strongly coupled to local operators, and therefore become
very fragile in the presence of external noise. We shall therefore restrict
ourselves to the models were no projective representations occur. In practice,
they are associated with the non-trivial elements of the group $H^{2}%
(G,U(1))$, and it turns out that for some interesting classes of groups such
as the dihedral groups $D_{N}$ with $N$ odd, $H^{2}(G,U(1))$ is
trivial~\cite{Propitius95}, so this restriction is not too important. As we
show in Section~\ref{sectiongenerators}, these assumptions allow us to find
all the possible phase factors associated with the gauge transformations in
terms of homomorphisms from the subgroups of $G$ that leave invariant a fixed
element of $G$ under conjugacy into $U(1)$. The last step is to construct a
set of local gauge-invariant operators corresponding to the following
elementary processes: moving a single fluxon, creating or annihilating a pair
of fluxons with vacuum quantum numbers, and branching processes were two
nearby fluxons merge into a single one (or conversely). We shall see that the
requirement of local gauge invariance leads to a relatively mild constraint
for the possible phase-factors, leaving a rather large set of solutions.

The main result of this work is twofold. First, we provide an explicit
computation of holonomy properties of fluxes in a set a models based on
dihedral groups. Of special interest is the simplest of them, $D_{3}$ which is
also the permutation group of 3 elements $S_{3}$. This group belongs to a
class which is in principle capable of universal quantum
computation~\cite{Mochon2004}. This part is therefore connected to the upper
layer (designing a set of univeral gates from properties of anyons) in the
classification outlined above. But our construction of a local Hamiltonian
version for a Chern-Simons model on a lattice provides some guidelines for
possible desirable physical implementations.

The plan of the paper is the following. Section II is mostly pedagogical,
providing the motivation for our general construction through the simplest
example of a Chern-Simons gauge theory, namely the non-compact $U(1)$ model.
In Section III we formulate general conditions on local Chern-Simons phase
factors that satisfy gauge invariance conditions. In Section IV we discuss the
construction of the electric field operator and derive condition on the phase
factor that allows one a gauge invariant fluxon dynamics. In Section V we
discuss the fluxon braiding properties and derive the Chern-Simons phase
factors associated with the braiding. In Section VI we apply our results to
the simplest non-Abelian groups $D_{n}$. Finally, Section VII gives
conclusions. Some technical points relevant for the general construction have
been relegated to Appendix A, and Appendix B discusses some interesting
features of the torus geometry. Although this geometry is not easy to
implement in experiments, it is conceptually very interesting since it is the
simplest example of a two-dimensional space without boundary and with
topologically non-trivial closed loops.

\section{Overview on Abelian Chern-Simons models}

To motivate the construction of the present paper, it is useful to discuss
some important features of Chern-Simons gauge theories in the simpler case of
Abelian groups. For this purpose, we shall consider here as an illustration
the model based on the \emph{continuous} Abelian group with one generator in
its non-compact version. Of course, our main purpose here is to address
\emph{finite} groups, but as we shall discuss, this non-compact $U(1)$ model
contains already the key ingredients. On a $2+1$ dimensional space-time, it is
defined from the following Lagrangian density:
\begin{equation}
\mathcal{L} = \frac{1}{2}\lambda(\dot{A}_{x}^{2} + \dot{A}_{y}^{2}) - \frac
{1}{2}\mu B^{2} + \nu(\dot{A}_{x}A_{y}-\dot{A}_{y}A_{x}) \label{L_CS}%
\end{equation}
where $B=\partial_{x}A_{y}-\partial_{y}A_{x}$ is the local magnetic field (a
pseudo-scalar in $2+1$ dimensions) and dots stand for time-derivatives. We
have used the gauge in which the time component $A_{0}$ of the vector
potential is zero. Because of this, we shall consider only invariance under
time-independent gauge transformations in this discussion. These are defined
as usual by \mbox{$A_{\rho}\rightarrow A_{\rho}+ \partial_{\rho}f$}, where
$f(x,y)$ is any time-independent smooth scalar function of spacial position.
Under such a gauge transformation, the action associated to the system
evolution on a two-dimensional space manifold $M$ during the time interval
$[t_{1},t_{2}]$ varies by the amount $\Delta\mathcal{S}$ where:
\begin{equation}
\label{defDeltaS}\Delta\mathcal{S}=\nu\int_{M} d^{2}\mathbf{r} \int_{t_{1}%
}^{t_{2}}dt\;\left(  \dot{A}_{x}\frac{\partial f}{\partial y} -\dot{A}%
_{y}\frac{\partial f}{\partial x}\right)
\end{equation}
Because $f$ is time-independent, the integrand is a total time-derivative, so
we may write \mbox{$\Delta \mathcal{S} = \nu (I(A_{2},f)-I(A_{1},f))$}, where
$A_{i}$ denotes the field configuration at time $t_{i}$, $i=1,2$, and:
\begin{equation}
\label{defI}I(A,f)=\int_{M} d^{2}\mathbf{r} \;\left(  A_{x}\frac{\partial
f}{\partial y} -A_{y}\frac{\partial f}{\partial x}\right)
\end{equation}
In the case where $M$ has no boundary (and in particular no hole), we may
integrate by parts and get:
\begin{equation}
I(A,f)=\int_{M} d^{2}\mathbf{r} \left(  \frac{\partial A_{y}}{\partial x}
-\frac{\partial A_{x}}{\partial y}\right)  f = \int_{M} d^{2}\mathbf{r} \;Bf
\label{Iwithoutboundary}%
\end{equation}
When $\nu\neq0$, this modifies in a non-trivial way the behavior of the
corresponding quantized model under time-independent gauge transformations.
One way to see this is to consider the time-evolution of the system's
wave-functional $\Psi(A,t)$. In a path-integral approach, the probability
amplitude to go from an initial configuration $A_{1}(\mathbf{r} )$ at time
$t_{1}$ to a final $A_{2}(\mathbf{r} )$ at time $t_{2}$ is given by:
\[
\mathcal{A}_{21}=\int\mathcal{D}A(\mathbf{r} ,t)\;\exp\{\frac{i}{\hbar
}\mathcal{S}(A)\}
\]
where fields $A(\mathbf{r} ,t)$ are required to satisfy boundary conditions
\mbox{$A(\rbf,t_{j})=A_{j}(\rbf)$} for $j=1,2$. After the gauge-transformation
\mbox{$\mathbf{A'}=\mathbf{A}+\mbox{\boldmath $\nabla$}f$}, and using the
above expression for $\Delta\mathcal{S}$, we see that the probability
amplitude connecting the transformed field configurations $A^{\prime}_{1}$ and
$A^{\prime}_{2}$ is:
\begin{equation}
\mathcal{A}_{2^{\prime}1^{\prime}}=\mathcal{A}_{21}\exp\left\{  i\frac{\nu
}{\hbar}(I(A_{2},f)-I(A_{1},f))\right\}  \label{gaugetransformedamplitude}%
\end{equation}
It is therefore natural to define the gauge-transformed wave-functional
$\tilde{\Psi}$ by:
\begin{equation}
\tilde{\Psi}(A^{\prime})=\Psi(A)\exp\left(  i\frac{\nu}{\hbar}I(A,f)\right)
\label{gaugetransformedwavefunctional}%
\end{equation}
This definition ensures that $\Psi(A,t)$ and $\tilde{\Psi}(A,t)$ evolve
according to the same set of probability amplitudes.

In a Hamiltonian formulation, we associate to any classical field
configuration $A(\mathbf{r} )$ a basis state $|A\rangle$. The
gauge-transformation corresponding to $f$ is now represented by a unitary
operator $U(f)$ defined by:
\begin{equation}
U(f)|A\rangle=\exp\left(  i\frac{\nu}{\hbar}I(A,f)\right)  |A^{\prime}%
\rangle\label{gaugetransformedbasisstate}%
\end{equation}
The presence of the phase-factor is one of the essential features of the
Chern-Simons term (i.e. the term proportional to $\nu$) added to the action.
Note that when $f$ varies, the family of operators $U(f)$ gives rise to a
representation of the full group of local gauge transformations. Indeed, at
the classical level, the composition law in this group is given by the
addition of the associated $f$ functions, and because $I(A,f)$ given in
Eq.~(\ref{Iwithoutboundary}) is itself gauge-invariant, we have
\mbox{$U(f)U(g)=U(f+g)$}. It is interesting to give an explicit expression for
$U(f)$. It reads:
\begin{align}
U(f)  &  = U_{\nu=0}(f)\exp\left\{  i\frac{\nu}{\hbar}\int_{M} d^{2}\mathbf{r}
\;Bf\right\} \label{explicitU(f)}\\
U_{\nu=0}(f)  &  = \exp\left\{  \frac{i}{\hbar}\int_{M} d^{2}\mathbf{r}
\;(\partial_{x} \Pi_{x}+ \partial_{y} \Pi_{y})f\right\} \nonumber
\end{align}
where $\Pi_{x}$ and $\Pi_{y}$ are the canonically conjugated variables to
$A_{x}$ and $A_{y}$. Note that this no longer holds in the case of a manifold
$M$ with a boundary, as will be discussed in a few paragraphs.

In the Hamiltonian quantization, a very important role is played by the
gauge-invariant electric operators $E_{x}$ and $E_{y}$. In the absence of
Chern-Simons term, they are simply equal to $\Pi_{x}$ and $\Pi_{y}$. When
$\nu\neq0$, because of Eq.~(\ref{explicitU(f)}), the transformation law for
$\Pi_{x}$ and $\Pi_{y}$ becomes:
\begin{align*}
\Pi_{x}  &  \rightarrow\Pi_{x} + \nu\partial_{y} f\\
\Pi_{y}  &  \rightarrow\Pi_{y} - \nu\partial_{x} f
\end{align*}
To compensate for this new gauge sensitivity of conjugated momenta, the
gauge-invariant electric field becomes:
\begin{align*}
E_{x}  &  = \Pi_{x} - \nu A_{y}\\
E_{y}  &  = \Pi_{y} + \nu A_{x}%
\end{align*}
Any classical gauge-invariant Lagrangian gives rise, after Legendre
transformation, to a Hamiltonian which is a functional of $E_{x}$, $E_{y}$,
and $B$ fields. If we add the Chern-Simons term to the original Lagrangian and
perform the Legendre transformation, we get a new Hamiltonian which is
expressed in terms of the new gauge-invariant $E_{x}$, $E_{y}$, and $B$ fields
through the \emph{same} functional as without the Chern-Simons term. For the
special example of the Maxwell-Chern-Simons Lagrangian~(\ref{L_CS}), this
functional reads:
\[
H=\int_{M} d^{2}\mathbf{r} \;\left(  \frac{1}{2\lambda}\mathbf{E}^{2}%
+\frac{\mu}{2}B^{2}\right)
\]
But although the Chern-Simons term preserves the Hamiltonian functional, it
does modify the dynamical properties of the system through a modification of
the basic commutation rules between $E_{x}$ and $E_{y}$. More precisely, we
have:
\begin{equation}
[E_{x}(\mathbf{r} ),E_{y}(\mathbf{r^{\prime}} )]=-i\hbar(2\nu)\delta
(\mathbf{r} -\mathbf{r^{\prime}} )
\end{equation}
So it appears that finding the appropriate deformations of electric field
operators plays a crucial role in constructing the Hamiltonian version of a
Chern-Simons theory. We have also seen that such deformations are strongly
constrained by the requirement of invariance under local gauge
transformations. This discussion shows that most of the relevant information
is implicitely encoded in the additional phase-factor $I(A,f)$ involved in
quantum gauge transformations, as in
Eqs.~(\ref{gaugetransformedwavefunctional}) and
(\ref{gaugetransformedbasisstate}).

Let us now briefly discuss what happens when the model is defined on a
two-dimensional space manifold $M$ with a boundary $\partial M$. Using Stokes'
formula, we may recast the phase factor $I(A,f)$ attached to a
gauge-transformation as:
\begin{equation}
I(A,f)=\int_{M} d^{2}\mathbf{r} \;Bf-\int_{\partial M}f\mathbf{A.dl}
\label{Iwithboundary}%
\end{equation}
In this situation, the phase factor $I(A,f)$ is no longer gauge-invariant, and
this implies that two gauge transformations attached to functions $f$ and $g$
do not commute because: \begin{widetext}
\begin{equation}
\left(I(A,f)+I(A+\nabla f,g)\right)-\left(I(A,g)+I(A+\nabla g,f)\right) =
\int_{\partial M}(f\mbox{\boldmath $\nabla$}g-g\mbox{\boldmath $\nabla$}f).\mathbf{dl}
\end{equation}
\end{widetext}
In more mathematical terms, this reflects the fact that the phase-factor
$I(A,f)$, as used in Eqs.~(\ref{gaugetransformedwavefunctional}) and
(\ref{gaugetransformedbasisstate}) defines only a projective representation of
the classical gauge group, that is:
\begin{equation}
U(f)U(g)=\exp\left(  -i\frac{\nu}{\hbar}\int_{\partial M}%
f\mbox{\boldmath $\nabla$}g.\mathbf{dl}\right)  U(f+g)
\end{equation}
As first shown by Witten~\cite{Witten89}, this may be understood in terms of a
chiral matter field attached to the boundary of $M$. An explicit example of
boundary degrees of freedom induced by a Chern-Simons term has been discussed
recently in the case of a
${\mathchoice {\hbox{$\sf\textstyle Z\kern-0.4em Z$}}
{\hbox{$\sf\textstyle Z\kern-0.4em Z$}} {\hbox{$\sf\scriptstyle
Z\kern-0.3em Z$}} {\hbox{$\sf\scriptscriptstyle Z\kern-0.2em Z$}}}_{2}$ model
on a triangular lattice~\cite{Doucot2005b}.

To close this preliminary section, it is useful to discuss the case of a
finite cylic group ${\mathchoice {\hbox{$\sf\textstyle Z\kern-0.4em Z$}}
{\hbox{$\sf\textstyle Z\kern-0.4em Z$}} {\hbox{$\sf\scriptstyle
Z\kern-0.3em Z$}} {\hbox{$\sf\scriptscriptstyle Z\kern-0.2em Z$}}}_{N}$. In
the $U(1)$ case, for a pair of points $\mathbf{r} $ and $\mathbf{r^{\prime}}
$, we have a natural group element defined by $\exp(i\frac{2\pi}{\Phi_{0}}%
\int_{\mathbf{r} }^{\mathbf{r^{\prime}} }\mathbf{A}.\mathbf{dl})$ where the
integral is taken along the segment joining $\mathbf{r} $ and
$\mathbf{r^{\prime}} $, and $\Phi_{0}$ is the flux quantum in the model. For a
finite group $G$, the notion of a Lie algebra is no longer available, so it is
natural to define the model on a lattice. In a classical gauge theory, each
oriented link $ij$ carries a group element $g_{ij} \in G$. We have the
important constraint $g_{ij}g_{ji}=e$, where $e$ is the neutral element of the
group $G$. In the quantum version, the Hilbert space attached to link $ij$ is
the finite dimensional space generated by the orthogonal basis $|g_{ij}%
\rangle$ where $g_{ij}$ runs over all elements of $G$. For a lattice, the
corresponding Hilbert space is obtained by taking the tensor product of all
these finite dimensional spaces associated to links. In the
${\mathchoice {\hbox{$\sf\textstyle Z\kern-0.4em Z$}}
{\hbox{$\sf\textstyle Z\kern-0.4em Z$}} {\hbox{$\sf\scriptstyle
Z\kern-0.3em Z$}} {\hbox{$\sf\scriptscriptstyle Z\kern-0.2em Z$}}}_{N}$ model,
$g_{ij}$ becomes an integer modulo $N$, $p_{ij}$. The connection with the
continuous case is obtained through the identification
\mbox{$\int_{i}^{j}\mathbf{A}.\mathbf{dl}=\frac{\Phi_0}{N}p_{ij}$}. On each
link $ij$, we introduce the unitary operator $\pi^{+}_{ij}$ which sends
$|p_{ij}\rangle$ into $|p_{ij}+1\rangle$. In the absence of a Chern-Simons
term, the generator of the gauge transformation based at site $i$ (which turns
$p_{jk}$ into $p_{jk}+\delta_{ji}-\delta_{ki}$) is $U_{i}=\prod_{j}^{(i)}%
\pi_{ij}^{+}$, where the product involves all the nearest neighbors of site
$i$. By analogy with the continuous case, the presence of a Chern-Simons term
is manifested by an additional phase-factor whose precise value depends on the
lattice geometry and is to some extent arbitrary, since fluxes are defined on
plaquettes, not on lattice sites. On a square lattice, a natural choice is to
define $U_{i}$ according to~\cite{Doucot2005}:
\begin{equation}
U_{i}=\prod_{j}^{(i)}\pi_{ij}^{+}\exp(-i\frac{\nu}{4\hbar}(\frac{2\pi}{N}%
)^{2}\sum_{(jk)\in\mathcal{L}(i)}p_{jk}) \label{DefUi}%
\end{equation}
where $\mathcal{L}(i)$ is the oriented loop defined by the outer boundary of
the four elementary plaquettes adjacent to site $i$. This expression has
exactly the same structure as Eq.~(\ref{explicitU(f)}), but somehow, the local
magnetic field at site $i$ is replaced by a smeared field on the smallest
available loop centered around $i$. It has been shown~\cite{Doucot2005} that a
consistent quantum theory can be constructed only when $\nu/\hbar$ is an
integer multiple of $N/\pi$.

\section{Generators of local gauge transformations}

\label{sectiongenerators}

As discussed in the previous section, the most important step is to construct
a non-trivial phase factor which appears in the definition of unitary
operators associated to local gauge transformations, generalizing
Eq.~(\ref{gaugetransformedbasisstate}). For this, let us first define the
operator $L_{ij}(g)$ which is the left multiplication of $g_{ij}$ by $g$,
namely: $L_{ij}(g)|g_{ij}\rangle=|gg_{ij}\rangle$. For any site $i$ and group
element $g$, we choose the generator of a local gauge transformation based at
$i$ to be of the following form:
\begin{equation}
\label{gaugegenerator}U_{i}(g)=\prod_{j}^{(i)}L_{ij}(g)\prod_{\mathbf{r}
}^{(i)}\chi_{\Phi(i,\mathbf{r} )}(g)
\end{equation}
where $j$ denotes any nearest neighbor of $i$ and $\Phi(i,\mathbf{r} )$ is the
flux around any of the four square plaquettes, centered at $\mathbf{r} $,
adjacent to $i$. Here, and in all this paper, we shall focus on the square
lattice geometry, to simplify the presentation. But adaptations of the basic
construction to other lattices are clearly possible. Since we are dealing with
a non-Abelian group, we have to specify an origin in order to define these
fluxes, and it is natural to choose the same site $i$, which is expressed
through the notation $\Phi(i,\mathbf{r} )$. Since we wish $U_{i}(g)$ to be
unitary, we require $|\chi_{\Phi}(g)|=1$. It is clear from this construction
that two generators $U_{i}(g)$ and $U_{j}(h)$ based on different sites
commute, since the phase factors $\chi_{\Phi}(g)$ are gauge invariant.

This form is a simple generalization of the lattice Chern-Simons models for
the cyclic groups ${\mathchoice {\hbox{$\sf\textstyle Z\kern-0.4em Z$}}
{\hbox{$\sf\textstyle Z\kern-0.4em Z$}} {\hbox{$\sf\scriptstyle
Z\kern-0.3em Z$}} {\hbox{$\sf\scriptscriptstyle Z\kern-0.2em Z$}}}_{N}$
discussed in the previous section. In this example, for a square plaquette
$ijkl$, the flux $\Phi$ is equal to $p_{ij}+p_{jk}+p_{kl}+p_{li}$ modulo $N$,
and $g$ is simply any integer modulo $N$. Eq.~(\ref{DefUi}) above may be
interpreted as:
\begin{equation}
\chi_{\Phi}(g)=\exp\left(  -i\frac{\nu}{4\hbar}(\frac{2\pi}{N})^{2}\Phi
g\right)
\end{equation}
This is a well defined function for $\Phi$ and $g$ modulo $N$ only if
$\nu/\hbar$ is an integer multiple of $2N/\pi$. We have not succeeded to cast
odd integer multiples of $N/\pi$ for $\nu/\hbar$ in the framework of the
general construction to be presented now. This is not too surprising since
these models were obtained by imposing special periodicity conditions on an
infinite-dimensional Hilbert space where $p_{ij}$ can take any integer
value~\cite{Doucot2005}. Our goal here is not to write down all possible
Chern-Simons theories with a finite group, but to easily construct a large
number of them, therefore allowing for non-trivial phase-factors when two
localized flux excitations are exchanged.

As discussed in the Introduction, a very desirable property, at least for the
sake of finding possible physical implementations, is that these deformed
generators define a unitary representation of the group $G$. So we wish to
impose:
\begin{equation}
U_{i}(g)U_{i}(h)=U_{i}(gh)
\end{equation}
or equivalently:
\begin{equation}
\chi_{h\Phi h^{-1}}(g)\chi_{\Phi}(h)=\chi_{\Phi}(gh)\label{defgaugegroup}%
\end{equation}
To solve these equations let us first choose a group element $\Phi$. Let us
denote by $H_{\Phi}$ the stabilizor of $\Phi$ under the operation of
conjugacy, namely $h$ belongs to $H_{\Phi}$ whenever
\mbox{$h\Phi h^{-1}=\Phi$}. This notion is useful to describe the elements of
the conjugacy class of $\Phi$. Indeed, we note that
\mbox{$gh\Phi (gh)^{-1}=g\Phi g^{-1}$} if $h$ belongs to $H_{\Phi}$.
Therefore, the elements in the conjugacy class of $\Phi$ are in one to one
correspondence with the left cosets of the form $gH_{\Phi}$. Let us pick one
representative $g_{n}$ in each of these cosets. We shall now find all the
functions \mbox{$\chi_{g_{n}\Phi g_{n}^{-1}}(g)$}. First we may specialize
Eq.~(\ref{defgaugegroup}) to the case where $h$ belongs to $H_{\Phi}$,
giving:
\begin{equation}
\chi_{\Phi}(g)\chi_{\Phi}(h)=\chi_{\Phi}(gh)
\end{equation}
In particular, it shows that the function \mbox{$h\rightarrow \chi_{\Phi}(h)$}
defines a group homomorphism from $H_{\Phi}$ to $U(1)$. Once this homomorphism
is known, we can specify completely $\chi_{\Phi}(g)$ for any group element $g$
once the values $\chi_{\Phi}(g_{n})$ for the coset representatives are known.
More explicitely, we have:
\begin{equation}
\chi_{\Phi}(g_{n}h)=\chi_{\Phi}(g_{n})\chi_{\Phi}(h)\label{defchi1}%
\end{equation}
where $h\in H_{\Phi}$. Finally, Eq.~(\ref{defgaugegroup}) yields:
\begin{equation}
\chi_{g_{n}\Phi g_{n}^{-1}}(g)=\frac{\chi_{\Phi}(gg_{n})}{\chi_{\Phi}(g_{n}%
)}\label{defchi2}%
\end{equation}
Let us now show that for any choice of homomorphism
\mbox{$h\rightarrow \chi_{\Phi}(h)$}, $h\in H_{\Phi}$, and unit complex
numbers for $\chi_{\Phi}(g_{n})$, Eqs.~(\ref{defchi1}), (\ref{defchi2})
reconstruct a function $\chi_{\Phi}(g)$ which satisfies the
condition~(\ref{defgaugegroup}). Any element $g^{\prime}$ in $G$ may be
written as $g^{\prime}=g_{n}h$, with $h\in H_{\Phi}$. We have:
\begin{align}
\chi_{g^{\prime}\Phi g^{\prime}{}^{-1}}(g) &  =\chi_{g_{n}\Phi g_{n}^{-1}%
}(g)\overset{(\ref{defchi2})}{=}\frac{\chi_{\Phi}(gg_{n})}{\chi_{\Phi}(g_{n}%
)}\overset{(\ref{defchi1})}{=}\nonumber\\
&  =\frac{\chi_{\Phi}(gg_{n}h)}{\chi_{\Phi}(g_{n}h)}=\frac{\chi_{\Phi
}(gg^{\prime})}{\chi_{\Phi}(g^{\prime})}%
\end{align}
which is exactly Eq.~(\ref{defgaugegroup}).

Note that there are many equivalent ways to choose these functions $\chi
_{\Phi}(g)$. Let us multiply the system wave-function by a phase-factor of the
form \mbox{$\prod_{i}\prod_{\rbf}^{(i)}\epsilon(\Phi(i,\rbf)),$} where
\mbox{$|\epsilon(\Phi)|=1$}. Under this unitary transformation, $\chi_{\Phi
}(g)$ is changed into $\tilde{\chi}_{\Phi}(g)$ given by:
\begin{equation}
\tilde{\chi}_{\Phi}(g)=\epsilon(g\Phi g^{-1})\chi_{\Phi}(g)\epsilon(\Phi)^{-1}%
\end{equation}
In particular, it is possible to choose the values of $\epsilon(g_{n}\Phi
g_{n}^{-1})$ so that $\tilde{\chi}_{\Phi}(g_{n})=1$. Although this does not
seem to be required at this stage of the construction, it is also necessary to
assume that when the flux $\Phi$ is equal to the neutral element $e$,
$\chi_{e}(g)=1$. This will play an important role later in ensuring that the
phase-factor accumulated by the system wave-function as a fluxon winds around
another is well defined.

\section{Basic processes for fluxon dynamics}

\label{sectionprocesses}

\subsection{General description}

\begin{figure}[th]
\includegraphics[width=2in]{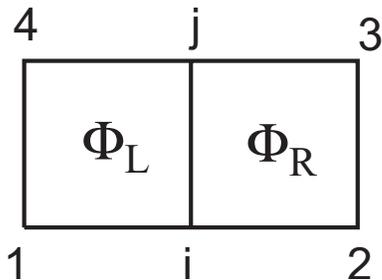} \caption{The site labeling
convention. For any bond $(ij)$ (shown here as middle vertical bond) the
surrounding sites are labeled $1,2,3,4$ as indicated here. The fluxes
$\Phi(i,L)$, $\Phi(i,R)$, ($\Phi(j,L)$, $\Phi(j,R)$) are counted
counterclockwise, starting from site $i$, ($j$), e.g. $\Phi(i,L)=g_{ij}%
g_{j4}g_{41}g_{1i}$ }%
\label{Twoplaquettes}%
\end{figure}

Our goal here is to construct local gauge invariant operations for basic
fluxon processes: fluxon motion, creation of a pair with vacuum quantum
numbers, branching of a single fluxon into two fluxons and their time
reversions. These three types of elementary processes can all be derived from
a single operation, the electric field operator that at the level of classical
configurations, is simply a left multiplication $L_{ij}(g)$ attached to any
link $ij$. To show this, let us consider a pair of two adjacent plaquettes as
shown on Fig.~\ref{Twoplaquettes}. We denote by $\Phi(i,L)$ (resp. $\Phi
(i,R)$) the local flux through the left (resp. right) plaquette, with site $i$
chosen as origin. Similarly, we define fluxes $\Phi(j,L)$ and $\Phi(j,R)$.
Changing the origin from $i$ to $j$ simply conjugates fluxes, according to
\mbox{$\Phi(j,L(R))=g_{ji}\Phi(i,L(R))g_{ij}$}. The left multiplication
$L_{ij}(g)$ changes $g_{ij}$ into \mbox{$g'_{ij}=gg_{ij}$}. Therefore, it
changes simultaneously both fluxes on the left and right plaquettes adjacent
to link $ij$. More specifically, we have:
\begin{align}
\Phi^{\prime}(i,L)  &  =g\Phi(i,L)\\
\Phi^{\prime}(i,R)  &  =\Phi(i,R)g^{-1}%
\end{align}
In particular, this implies:
\begin{align}
\Phi^{\prime}(i,R)\Phi^{\prime}(i,L)  &  =\Phi(i,R)\Phi(i,L)\\
\Phi^{\prime}(i,L)\Phi^{\prime}(i,R)  &  =g\Phi(i,L)\Phi(i,R)g^{-1}%
\end{align}
Note that transformation laws for fluxes based at site $j$ are slightly more
complicated since they read:
\begin{align}
\Phi^{\prime}(j,L)  &  =\Phi(j,L)(g_{ji}gg_{ij})\\
\Phi^{\prime}(j,R)  &  =(g_{ji}g^{-1}g_{ij})\Phi(j,R)
\end{align}
This asymmetry between $i$ and $j$ arises because
\mbox{$g'_{ji}=g_{ji}g^{-1}$}, so we have:
\begin{equation}
L_{ij}(g)=R_{ji}(g^{-1})
\end{equation}
where $R_{ji}(h)$ denotes the right multiplication of $g_{ji}$ by the group
element $h$. In the absence of Chern-Simons term, $L_{ij}(g)$ commutes with
all local gauge generators with the exception of $U_{i}(h)$ since:
\begin{equation}
U_{i}(h)L_{ij}(g)U_{i}(h)^{-1}=L_{ij}(hgh^{-1})
\end{equation}

We now apply these general formulas to the elementary processes involving
fluxes. Suppose that initially a flux $\Phi$ was localized on the left
plaquette, and that the right plaquette is fluxless. Applying $L_{ij}%
(g=\Phi^{-1})$ on such configuration gives:
\begin{align}
\Phi^{\prime}(i,L)  &  =e\\
\Phi^{\prime}(i,R)  &  =\Phi
\end{align}
which shows that a $\Phi$-fluxon has moved from the left to the right
plaquette. A second interesting situation occurs when both plaquettes are
initially fluxless. The action of $L_{ij}(\Phi^{-1})$ on such state produces a
new configuration characterized by:
\begin{align}
\Phi^{\prime}(i,L)  &  =\Phi^{-1}\\
\Phi^{\prime}(i,R)  &  =\Phi
\end{align}
So we have simply created a fluxon and antifluxon pair from the vacuum. Of
course, applying $L_{ij}(\Phi)$ on the final state annihilates this pair.
Finally, a single flux \mbox{$\Phi=\Phi_{2}\Phi_{1}$} originaly located on the
left plaquette may split into a pair $\Phi_{1}$ on the left and $\Phi_{2}$ on
the right. This is achieved simply by applying $L_{ij}(\Phi_{2}^{-1})$.

In order to incorporate these elementary processes into a Hamiltonian
Chern-Simons theory, we have to modify $L_{ij}(g)$ into an electric field
operator $\mathcal{E}_{ij}(g)$ by introducing phase-factors so that it
commutes for all generators $U_{k}(h)$ if $k\neq i$ and that:
\begin{equation}
U_{i}(h)\mathcal{E}_{ij}(g)U_{i}(h)^{-1}=\mathcal{E}_{ij}(hgh^{-1})
\label{fieldselfcharge}%
\end{equation}
As explained in the introduction, we shall need only to construct
$\mathcal{E}_{ij}(g)$ for special types of configurations for which at least
one of the four fluxes $\Phi(i,L),\Phi(i,R),\Phi^{\prime}(i,L),\Phi^{\prime
}(i,R)$ vanishes. Nevertheless, it is useful to first construct $\mathcal{E}%
_{ij}(g)$ for an arbitrary flux background. The requirement of local
gauge-symmetry induces strong constraints on the phase-factors $\chi_{\Phi
}(h)$ as we shall now show. These constraints are less stringent when we
restrict the action of $\mathcal{E}_{ij}(g)$ to the limited set of
configurations just discussed.

\subsection{Construction of gauge-invariant electric field operators}

\begin{figure}[th]
\includegraphics[width=3.0in]{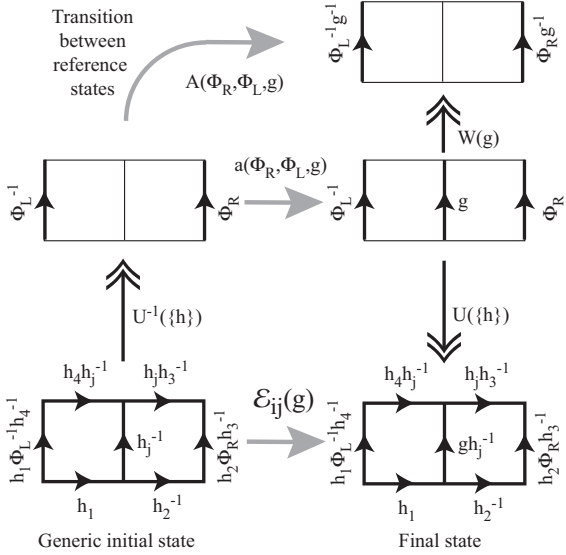}
\caption{Construction
of the gauge-invariant electric field operator $\mathcal{E}_{ij}(g)$. This
operator transforms the pair of fluxes ($\Phi_{L}$, $\Phi_{R}$) of the left
and right plaquettes into the pair ($g\Phi_{L}$, $\Phi_{R}g^{-1}$). We define
the amplitude of the transition between two reference states shown at the top
of the figure by $A(\Phi_{L},\Phi_{R},g)$. The amplitude of the process
starting from a generic initial state shown in the lower left can be related
to $A(\Phi_{L},\Phi_{R},g)$ using the gauge invariance. This is done in two
steps. First, the gauge transformation
\mbox{$U(\{h\})=U_{1}(h_{1})U_{2}(h_{2})U_{3}(h_{3})U_{4}(h_{4})
U_{j}(h_{j})$} is used to relate the amplitude of the process starting from
the generic state to the amplitude of the transition $a(\Phi_{L},\Phi_{R},g)$
between reference state and a special state shown in the middle right. Second,
we use gauge transformation $W(g)=U_{4}(g)U_{j}(g)U_{3}(g)$ to relate the
special state to the reference state on the upper right. Site labeling is the
same as in Figure 1. }%
\label{ElectricFieldConstruction}%
\end{figure}

We now construct the electric field operators $\mathcal{E}_{ij}(g)$ attached
to links. In the absence of Chern-Simons term the electric field operator is
equivalent to the left multiplication of the link variable:
\begin{equation}
\mathcal{E}_{ij}(g)=L_{ij}(g)
\end{equation}
In this case, $\mathcal{E}_{ij}(g)$ commutes with all local gauge generators
with the exception of $U_{i}(h)$, and Eq.~(\ref{fieldselfcharge}) is
satisfied. We have also the group property, namely:
\begin{equation}
\mathcal{E}_{ij}(g)\mathcal{E}_{ij}(h)=\mathcal{E}_{ij}(gh)
\label{electricgroup}%
\end{equation}

The Chern-Simons term gives phase factors $\chi_{\Phi}(g)$ to the gauge
generators, so we expect some phase factor, $\Upsilon_{ij},$ to appear in the
electric field operators as well: $\mathcal{E}=L_{ij}\Upsilon_{ij}$. The gauge
invariance condition allows us to relate the phase factors associated with
different field configurations to each other. Specifically, we introduce the
reference states (shown on the top of Fig. \ref{ElectricFieldConstruction} )
in which only two bonds carry non-trivial group elements. We define the
transition amplitude induced by the electric field between these reference
states by $A(\Phi_{L},\Phi_{R},g)$. In order to find the phases $\Upsilon
_{ij}$ for arbitrary field configuration we first transform both the initial
and final state by $U(\{h\})=U_{1}(h_{1})U_{2}(h_{2})U_{3}(h_{3})U_{4}%
(h_{4})U_{j}(h_{j})$. This relates the amplitude of the generic process to the
amplitude, $a(\Phi_{L},\Phi_{R},g)$, of the process that starts with the
reference state but leads to the special final state shown in Fig.
\ref{ElectricFieldConstruction} middle right. Collecting the phase factors
associated with the gauge transformation $U(\{h\})$ before and after the
electric field moved the flux we get%
\begin{align}
\Upsilon_{ij}  &  =\frac{\chi_{\Phi^{\prime}(i,L)}(h_{1})}{\chi_{\Phi
(i,L)}(h_{1})}\frac{\chi_{\Phi^{\prime}(i,R)}(h_{2})}{\chi_{\Phi(i,R)}(h_{2}%
)}\frac{\chi_{\Phi^{\prime}(j,R)}(h_{3})}{\chi_{\Phi(j,R)}(h_{3})}\nonumber\\
&  \times\frac{\chi_{\Phi^{\prime}(j,L)}(h_{4})}{\chi_{\Phi(j,L)}(h_{4})}%
\frac{\chi_{\Phi^{\prime}(j,L)}(h_{j})}{\chi_{\Phi(j,L)}(h_{j})} \frac
{\chi_{\Phi^{\prime}(j,R)}(h_{j})}{\chi_{\Phi(j,R)}(h_{j})}\nonumber\\
&  \times a(\Phi_{L},\Phi_{R},g)
\end{align}
where $\Phi(i,L)$ denotes the flux in the left plaquette counted from site
$i,$ $\Phi(j,R)$ denotes flux in the right plaquette counted from site $j$,
and prime refers to the flux configuration after the action of the electric
field. Finally, we employ the gauge transformation $W(g)=U_{4}(g)U_{j}%
(g)U_{3}(g)$ to relate this special final state to the reference state. The
phase factor associated with this gauge transformation is $\left(  \chi
_{\Phi_{L}g}(g)\chi_{g^{-1}\Phi_{R}}(g)\right)  ^{2}$ so
\begin{equation}
a(\Phi_{L},\Phi_{R},g)=\frac{A(\Phi_{L},\Phi_{R},g)}{\left(  \chi_{\Phi_{L}%
g}(g)\chi_{g^{-1}\Phi_{R}}(g)\right)  ^{2}}%
\end{equation}
In order to express the phase factors, $\Upsilon_{ij}$, through the initial
field configuration we relate the parameters, $h_{k}$, of the gauge
transformations to the bond variables $g_{k}$ by
\begin{align*}
h_{1}  &  =g_{1i},\;\;\;\;h_{2}=g_{2i},\;\;\;\;h_{3}=g_{3j}g_{ji},\\
h_{4}  &  =g_{4j}g_{ji},\;\;\;\;h_{j}=g_{ji}%
\end{align*}
and the fluxes in the left and right plaquettes before and after electric
field operator has changed them. Before the electric field operator has acted
the fluxes were
\begin{align*}
\Phi(i,L)  &  =\Phi_{L}\\
\Phi(j,L)  &  =g_{ji}\Phi_{L}g_{ij}\\
\Phi(i,R)  &  =\Phi_{R}\\
\Phi(j,R)  &  =g_{ji}\Phi_{R}g_{ij}%
\end{align*}
while afterwards they become
\begin{align*}
\Phi^{\prime}(i,L)  &  =g\Phi_{L}\\
\Phi^{\prime}(j,L)  &  =g_{ji}\Phi_{L}gg_{ij}\\
\Phi^{\prime}(i,R)  &  =\Phi_{R}g^{-1}\\
\Phi^{\prime}(j,R)  &  =g_{ji}g^{-1}\Phi_{R}g_{ij}%
\end{align*}

Combining the preceding equations and using the relation~(\ref{defgaugegroup})
a few times, we get the final expression for the phase-factor,
\begin{align}
\Upsilon_{ij}  &  =\frac{\chi_{\Phi^{\prime}(i,L)}(g_{1i})}{\chi_{\Phi
(i,L)}(g_{1i})}\frac{\chi_{\Phi^{\prime}(i,R)}(g_{2i})}{\chi_{\Phi
(i,R)}(g_{2i})}\frac{\chi_{\Phi^{\prime}(j,L)}(g_{4j})}{\chi_{\Phi
(j,L)}(g_{4j})}\nonumber\\
&  \times\frac{\chi_{\Phi^{\prime}(j,R)}(g_{3j})}{\chi_{\Phi(j,R)}(g_{3j}%
)}\frac{\chi_{\Phi^{\prime}(i,L)}(g_{ji}^{\prime})^{2}}{\chi_{\Phi
(i,L)}(g_{ji})^{2}}\frac{\chi_{\Phi^{\prime}(i,R)}(g_{ji}^{\prime})^{2}}%
{\chi_{\Phi(i,R)}(g_{ji})^{2}}\nonumber\\
&  \times A(\Phi_{L},\Phi_{R},g) \label{ElPhaseFactor}%
\end{align}
where we have used the definition $g_{ij}^{\prime}=gg_{ij}$
(\mbox{$g_{ji}^{\prime}=g_{ji}g^{-1}$}) to make it more symmetric.

Commutation of $\mathcal{E}_{ij}(g)$ with $U_{1}(h_{1})$, $U_{2}(h_{2})$,
$U_{3}(h_{3})$, $U_{4}(h_{4})$, and $U_{j}(h_{j})$ follows directly from this
construction. It can also be checked directly from~(\ref{ElPhaseFactor}),
using the condition~(\ref{defgaugegroup}) on the elementary phase-factors
$\chi_{\Phi}(g)$. Note that sites $i$ and $j$ play different roles, which is
expected because $\mathcal{E}_{ij}(g)$ acts by \emph{left} multiplication on
$g_{ij}$ whereas $\mathcal{E}_{ji}(g)$ acts by \emph{right} multiplication on
the same quantity.

\begin{figure}[th]
\includegraphics[width=3.0in]{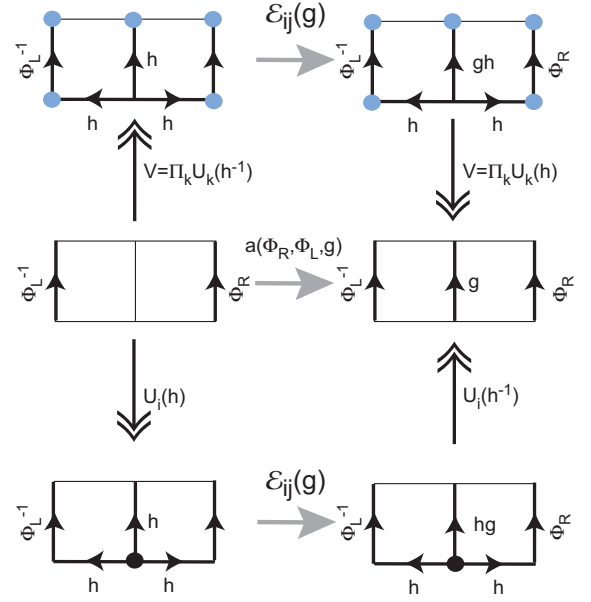}
\caption{(Color
online) Gauge invariance of electric field operator implies condition on the
phase factors $\chi_{\Phi}(g)$. The state shown at the top and botton left of
the figure can be transformed to the reference state (middle) in two ways: by
applying a gauge transformation $V=\prod_{k\neq i}U_{k}(h)$ on all shaded
sites or by applying transformation $U_{i}(h^{-1})$ on site $i$. If $\Phi
_{L}h=h\Phi_{L}$ and $\Phi_{R}h=h\Phi_{R}$ these transformations lead to the
same reference state shown in the middle. Furthermore, the action of electric
field on these states can be found by making a gauge transformation of the
final state shown on the middle right if $gh=hg$. The phases of electric field
operator obtained in these two ways should be equal giving additional
condition on the phase associated with the gauge field transformation.}%
\label{ElectricFieldGaugeInvariance}%
\end{figure}

\bigskip

Although the electric field operator commutes with $U_{1}(h_{1})$,
$U_{2}(h_{2})$, $U_{3}(h_{3})$, $U_{4}(h_{4})$, and $U_{j}(h_{j})$, it does
not necessarily commute with $U_{i}(h)$ even if $hgh^{-1}=g$. In fact, the
requirement of this commutation leads to an important constraint on possible
choices of phases $\chi_{\Phi}(g)$. The appearance of the new constraints
becomes clear when one considers a special field configuration shown in Fig.
\ref{ElectricFieldGaugeInvariance}. Two identical field configurations shown
on the top and bottom left of this figure can be obtained by two different
gauge transformations from the reference state if both $\Phi_{L}$ and
$\Phi_{R}$ commute with $h$: in one case one applies gauge transformation
\emph{only} at site $i$ , in the other one applies gauge transformation on all
sites \emph{except} $i$. Provided that the resulting states are the same, i.e.
$gh=hg$, the total phase factor $\Upsilon_{ij}$ obtained by these two
different ways should be the same.

The phase factors associated with these gauge-transformations are the
following
\begin{align*}
U_{i}(h)  &  \!\rightarrow\!\chi_{\Phi_{L}}(h)\chi_{\Phi_{R}}(h)\\
U_{i}^{-1}(h)  &  \!\rightarrow\!\frac{1}{\chi_{g\Phi_{L}}(h)\chi_{\Phi
_{R}g^{-1}}(h)}\\
V_{i}^{-1}(h)  &  \!\rightarrow\!\chi_{\Phi_{L}}^{3}(h^{-1})\chi_{\Phi_{R}%
}^{3}(h^{-1})\\
V_{i}(h)  &  \!\rightarrow\!\frac{1}{\chi_{\Phi_{L}g}^{2}(\!h^{-1}%
\!)\chi_{g\Phi_{L}}(\!h^{-1}\!)\chi_{g^{-1}\Phi_{R}}^{2}(\!h^{-1}\!)\chi
_{\Phi_{R}g^{-1}}(\!h^{-1}\!)}%
\end{align*}
Putting all these factors together we conclude that the gauge invariance of
the electric field operator implies that
\begin{align}
&  \frac{\chi_{\Phi_{L}}(h)\chi_{\Phi_{R}}(h)}{\chi_{g\Phi_{L}}(h)\chi
_{\Phi_{R}g^{-1}}(h)}=\nonumber\\
&  \frac{\chi_{\Phi_{L}}^{3}(h^{-1})\chi_{\Phi_{R}}^{3}(h^{-1})}{\chi
_{\Phi_{L}g}^{2}(\!h^{-1}\!)\chi_{g\Phi_{L}}(\!h^{-1}\!)\chi_{g^{-1}\Phi_{R}%
}^{2}(\!h^{-1}\!)\chi_{\Phi_{R}g^{-1}}(\!h^{-1}\!)}%
\end{align}
This condition can be further simplified by using the main phase factor
equation (\ref{defgaugegroup}). We start by noting that because $h$ commutes
with $\Phi_{L}$, $\Phi_{R}$, and $g$, $\chi_{\Phi_{R,L}}(h)\chi_{\Phi_{R,L}%
}(h^{-1})\equiv1$ and $\chi_{g\Phi_{R,L}}(h)\chi_{g\Phi_{R,L}}(h^{-1})\equiv
1$. This gives
\begin{equation}
\chi_{\Phi_{L}}^{4}(h)\chi_{\Phi_{R}}^{4}(h)=\chi_{\Phi_{L}g}^{2}%
(h)\chi_{g\Phi_{L}}^{2}(h)\chi_{g^{-1}\Phi_{R}}^{2}(h)\chi_{\Phi_{R}g^{-1}%
}^{2}(h) \label{almostrelation}%
\end{equation}
Furthermore, combining the identities
\begin{align*}
\chi_{g\Phi_{L}}(h)  &  =\chi_{g\Phi_{L}}(ghg^{-1})\\
&  =\chi_{\Phi_{L}g}(g)\chi_{\Phi_{L}g}(h)\chi_{g\Phi_{L}}(g^{-1})
\end{align*}
and%
\[
1=\chi_{g\Phi_{L}}(gg^{-1})=\chi_{\Phi_{L}g}(g)\chi_{g\Phi_{L}}(g^{-1})
\]
we get
\begin{align}
\chi_{g\Phi_{L}}(h)  &  =\chi_{\Phi_{L}g}(h)\\
\chi_{\Phi_{R}g^{-1}}(h)  &  =\chi_{g^{-1}\Phi_{R}}(h)
\end{align}
This reduces the condition~(\ref{almostrelation}) to a much simpler final form%

\begin{equation}
\left(  \frac{\chi_{g\Phi_{L}}(h)}{\chi_{\Phi_{L}}(h)}\frac{\chi_{\Phi
_{R}g^{-1}}(h)}{\chi_{\Phi_{R}}(h)}\right)  ^{4}=1
\label{PhaseFactorCondition}%
\end{equation}
We emphasize that constraint (\ref{PhaseFactorCondition}) on the phase factors
has to be satisfied only for fluxes satisfying the condition $(h\Phi_{L}%
h^{-1},h\Phi_{R}h^{-1},hgh^{-1})=(\Phi_{L},\Phi_{R},g)$. Although we have
derived this condition imposing only the gauge invariance of the electrical
field acting on a very special field configuration, a more lengthy analysis
shows that it is sufficient to ensure that in a general case%
\begin{equation}
U_{i}(h)\mathcal{E}_{ij}(g)|\Psi\rangle=\mathcal{E}_{ij}(hgh^{-1}%
)U_{i}(h)|\Psi\rangle
\end{equation}
The details of the proof are presented in Appendix ~\ref{amplsol}.

Unlike Eq.~(\ref{defgaugegroup}), the constraint (\ref{PhaseFactorCondition})
relates functions $\chi_{\Phi}(g)$ and $\chi_{\Phi^{\prime}}(g)$ for $\Phi$
and $\Phi^{\prime}$ belonging to \emph{different} conjugacy classes. As shown
in section~\ref{classmodels} this constraint strongly reduces the number of
possible Chern-Simons theories. Note that if both $\chi_{\Phi}^{(1)}(g)$ and
$\chi_{\Phi}^{(2)}(g)$ are solutions of the fundamental
relations~(\ref{defgaugegroup}) and~(\ref{PhaseFactorCondition}), their
product $\chi_{\Phi}^{(1)}(g)\chi_{\Phi}^{(2)}(g)$ is also a solution. So
there is a natural group structure on the set of possible Chern-Simons models
based on the group $G$, which is transparent in the path-integral description:
this means that the sum of two Chern-Simons action is also a valid
Chern-Simons action.

Is this construction also compatible with the group
property~(\ref{electricgroup})? From Eq.~(\ref{PhaseFactorCondition}), we
obtain:
\begin{equation}
\mathcal{E}_{ij}(g^{\prime})\mathcal{E}_{ij}(g)|\Psi\rangle=\beta(\Phi
_{L},\Phi_{R},g)\mathcal{E}_{ij}(g^{\prime}g)|\Psi\rangle
\end{equation}
where:
\[
\beta(\Phi_{L},\Phi_{R},g)=\frac{A(\Phi_{L},\Phi_{R},g)A(g\Phi_{L},\Phi
_{R}g^{-1},g^{\prime})}{A(\Phi_{L},\Phi_{R},g^{\prime}g)}%
\]
It does not seem that $\beta(\Phi_{L},\Phi_{R},g)$ are always equal to unity
for any choice of $\chi_{\Phi}(g)$ that in turns determines the amplitudes
$A(s)$ (see Appendix~\ref{amplsol}). But this is not really a problem because
this group property does not play much role in the construction of
gauge-invariant Hamiltonians.

We now specialize the most general constraint arising from local
gauge-invariance to the various physical processes which are required for
fluxon dynamics. For the single fluxon moving operation, we have
\mbox{$\Phi_{L}=\Phi$}, \mbox{$\Phi_{R}=e$}, and \mbox{$g=\Phi^{-1}$}, so the
condition Eq.~(\ref{PhaseFactorCondition}) is always satisfied. For the pair
creation process, we have \mbox{$\Phi_{L}=\Phi_{R}=e$}, and
\mbox{$g=\Phi^{-1}$}. The constraint becomes:
\begin{equation}
\left(  \chi_{\Phi^{-1}}(h)\chi_{\Phi}(h)\right)  ^{4}=1\;\;\mathrm{if}%
\;\;h\Phi h^{-1}=\Phi\label{vacuumpaircond}%
\end{equation}
Finally, let us consider the splitting of an isolated fluxon into two nearby
ones. This is described by \mbox{$\Phi_{L}=\Phi_{2}\Phi_{1}$},
\mbox{$\Phi_{R}=e$}, and \mbox{$g=\Phi_{2}^{-1}$}. We need then to impose:
\begin{equation}
\left(  \frac{\chi_{\Phi_{1}}(h)\chi_{\Phi_{2}}(h)}{\chi_{\Phi_{1}\Phi_{2}%
}(h)}\right)  ^{4}=1\;\mathrm{if}\;(h\Phi_{1}h^{-1},h\Phi_{2}h^{-1})=(\Phi
_{1},\Phi_{2}) \label{generalpaircond}%
\end{equation}
It is clear that condition~(\ref{vacuumpaircond}) is a special case of the
stronger condition~(\ref{generalpaircond}). Furthermore, multiplying the
conditions (\ref{generalpaircond}) for pairs of fluxes $(\Phi_{L},\Phi_{R})$
and $(g\Phi_{L},\Phi_{R}g^{-1})$ we get the most general
condition~(\ref{PhaseFactorCondition}); this shows that constraint
(\ref{generalpaircond}) is necessary and sufficient condition for the gauge
invariant definition of electric field operator acting on any flux configuration.

\section{General expression for holonomy of fluxons}

\begin{figure}[th]
\includegraphics[width=2.0in]{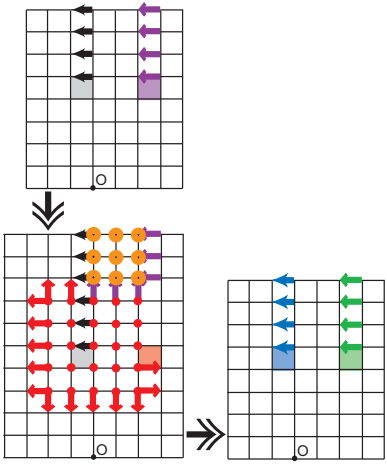}\caption{(Color online) Braiding of
two fluxes yields in a non-trivial transformation of their values and a phase
factor. We start with the flux configuration shown in upper pane with fluxes
$g_{1}$ (right) and $g_{2}$ (left) connected to the edge by the strings of
$g_{1}$ (purple arrow) or $g_{2}$ (black) group elements. Moving the right
flux around the left leaves behind a string of $g_{1}$ elements until its path
crosses the vertical string of $g_{2}$. Upon crossing the string changes to
the string of $g_{1}^{\prime}=g_{2}^{-1}g_{1}g_{2}$ (red arrows). Performing
the gauge transformations with $h=(g_{1}^{\prime})$ on all sites indicated by
full (red) dots and with $h=g_{1}h$ on sites indicated by empty (orange) dots
reduces the configuration to the template shown in the last pane with the new
fluxes $g_{1}^{\prime}$ and $g_{2}^{\prime}$ (see text, Eq.~(\ref{Braiding})).
}%
\label{FluxBraiding}%
\end{figure}

Let us consider two isolated fluxes carrying group elements $g_{1}$ and
$g_{2}$, and move the first one counterclockwise around the second one, as
shown on Fig.~\ref{FluxBraiding}. This can be done by successive applications
of local gauge-invariant electric field operators as discussed in the previous
section. Although we wish to work in the gauge-invariant subspace, it is very
convenient to use special configurations of link variables to illustrate what
happens in such a process. We simply have to project all special states on the
gauge invariant subspace, which is straightforward since the fluxon moving
operator commutes with this projector. The initial fluxes are described by two
vertical strings of links carrying the group elements $g_{1}$ and $g_{2}$, see
Fig.~\ref{FluxBraiding}. When several other fluxes are present, besides the
two to be exchanged, it is necessary to choose the location of the strings in
such a way that no other flux is present in the vertical band delimited by the
strings attached to the two fluxons. During the exchange process, the first
fluxon collides with the second string, and after this event, it no longer
carries the group element $g_{1}$, but its conjugate
\mbox{$g'_{1}=g_{2}^{-1}g_{1}g_{2}$}. After the process is completed, the
first flux has recovered its original position, but the configuration of group
elements has changed. If we measure the second flux using a path starting at
point O shown on Fig.~\ref{FluxBraiding} we find that it has also changed into
\mbox{$g'_{2}=g_{2}^{-1}g_{1}^{-1}g_{2}g_{1}g_{2}$}. The final state can be
reduced to its template state built from two strings carrying groups elements
$g^{\prime}_{1}$ and $g^{\prime}_{2}$ by the gauge transformation $\prod_{i}
U_{i}(h_{i})$ where $h_{i}$ is locally constant in the two following regions:
the core region inside the circuit followed by the first fluxon ($h_{i}%
=h_{\mathrm{core}}$), the intermediate region delimited by the two initial
vertical strings and the upper part of the circuit ($h_{i}=h_{\mathrm{int}}$).
Note that because we do not wish to modify external fluxes, we cannot perform
gauge transformations in the bulk, outside of these regions.

Group elements $h_{\mathrm{core}}$ and $h_{\mathrm{int}}$ have to satisfy the
following conditions, which may be obtained readily upon inspection of
Fig.~\ref{FluxBraiding}
\begin{align*}
h_{\mathrm{core}}  &  =g_{1}^{\prime}\\
h_{\mathrm{core}}g_{1}^{\prime-1}  &  =e\\
h_{\mathrm{int}}g_{2}^{\prime}  &  =g_{2}\\
h_{\mathrm{core}}h_{\mathrm{int}}^{-1}  &  =g_{1}\\
g_{1}^{\prime}h_{\mathrm{int}}^{-1}  &  =g_{1}\\
h_{\mathrm{core}}g_{2}^{\prime}h_{\mathrm{core}}^{-1}  &  =g_{2}%
\end{align*}
These equations are mutually compatible, and we get
\mbox{$h_{\mathrm{int}}=g_{1}^{-1}g_{2}^{-1}g_{1}g_{2}$}. Since fluxes are
present in the core region, they will contribute a phase-factor $f(g_{1}%
,g_{2})$ when the gauge transformation from the template to the actual final
state is performed. The final result may be summarized as follows:
\begin{align}
|g_{1},g_{2}\rangle &  \longrightarrow f(g_{1},g_{2})|g_{1}^{\prime}%
,g_{2}^{\prime}\rangle\\
g_{1}^{\prime}  &  =g_{2}^{-1}g_{1}g_{2}\\
g_{2}^{\prime}  &  =g_{2}^{-1}g_{1}^{-1}g_{2}g_{1}g_{2}\label{Braiding}\\
f(g_{1},g_{2})  &  =\chi_{g_{1}^{\prime}}(g_{1}^{\prime})^{2}\chi
_{g_{2}^{\prime}}(g_{1}^{\prime})^{4} \label{adiabaticphase}%
\end{align}
The new phase-factor $f(g_{1},g_{2})$ appears not to depend on the detailed
path taken by the first fluxon, but just on the fact that it winds exactly
once around the second. In this sense, our construction really implements a
topological theory.

\section{Application to dihedral groups}

\subsection{General properties of dihedral groups}

\label{genpropdihedral}

The dihedral groups $D_{N}$ are among the most natural to consider, since they
contain a normal cyclic group
${\mathchoice {\hbox{$\sf\textstyle Z\kern-0.4em Z$}}
{\hbox{$\sf\textstyle Z\kern-0.4em Z$}} {\hbox{$\sf\scriptstyle
Z\kern-0.3em Z$}} {\hbox{$\sf\scriptscriptstyle Z\kern-0.2em Z$}}}_{N}$ which
is simply extended by a ${\mathchoice {\hbox{$\sf\textstyle Z\kern-0.4em Z$}}
{\hbox{$\sf\textstyle Z\kern-0.4em Z$}} {\hbox{$\sf\scriptstyle
Z\kern-0.3em Z$}} {\hbox{$\sf\scriptscriptstyle Z\kern-0.2em Z$}}}_{2}$ factor
to form a semi-direct product. In this respect, one may view this family as
the most weakly non-Abelian groups. $D_{N}$ can be described as the isometry
group of a planar regular polygon with $N$ vertices. The
${\mathchoice {\hbox{$\sf\textstyle Z\kern-0.4em Z$}}
{\hbox{$\sf\textstyle Z\kern-0.4em Z$}} {\hbox{$\sf\scriptstyle
Z\kern-0.3em Z$}} {\hbox{$\sf\scriptscriptstyle Z\kern-0.2em Z$}}}_{N}$
subgroup corresponds to rotations with angles multiples of $2\pi/N$. We shall
denote by $\mathcal{C}$ the generator of this subgroup, so $\mathcal{C}$ may
be identified with the $2\pi/N$ rotation. $D_{N}$ contains also $N$
reflections, of the form $\tau\mathcal{C}^{n}$. The two elements $\mathcal{C}$
and $\tau$ generate $D_{N}$, and they are subjected to the following minimal
set of relations:
\begin{align}
\mathcal{C}^{N}  &  = e\label{genrel1}\\
\tau^{2}  &  = e\label{genrel2}\\
\tau\mathcal{C} \tau &  = \mathcal{C}^{-1} \label{genrel3}%
\end{align}
This last relation shows that indeed $D_{N}$ is non-Abelian.

The next useful information about these groups is the list of conjugacy
classes. If $N=2M+1$, $D_{N}$ contains $M+2$ classes which are:
\mbox{$\{e\}$}, \mbox{$\{\mathcal{C},\mathcal{C}^{-1}\}$},...,
\mbox{$\{\mathcal{C}^{M},\mathcal{C}^{-M}\}$},
\mbox{$\{\tau,\tau\mathcal{C},...,\tau\mathcal{C}^{N-1}\}$}. If $N=2M$, there
are $M+3$ classes whose list is: \mbox{$\{e\}$}, \mbox{$\{\mathcal{C}^{M}\}$},
\mbox{$\{\mathcal{C},\mathcal{C}^{-1}\}$},...,
\mbox{$\{\mathcal{C}^{M-1},\mathcal{C}^{-M+1}\}$},
\mbox{$\{\tau,\tau\mathcal{C}^{2},...,\tau\mathcal{C}^{N-2}\}$}, \mbox{$\{\tau\mathcal{C},\tau\mathcal{C}^{3},...,\tau\mathcal{C}^{N-1}\}$}.

As shown in section~\ref{sectiongenerators}, in order to construct possible
phase factors associated to gauge transformations, we need to know the
stabilizors of group elements for the conjugacy operation of $D_{N}$ acting on
itself. Here is a list of these stabilizors:\newline For $N$ odd:
\begin{align*}
e  &  \rightarrow D_{N}\\
\mathcal{C}^{p}  &  \rightarrow
{\mathchoice {\hbox{$\sf\textstyle Z\kern-0.4em Z$}} {\hbox{$\sf\textstyle Z\kern-0.4em Z$}} {\hbox{$\sf\scriptstyle
Z\kern-0.3em Z$}} {\hbox{$\sf\scriptscriptstyle Z\kern-0.2em Z$}}}%
_{N},\;\;\;1\le p \le N-1\\
\tau\mathcal{C}^{p}  &  \rightarrow\{e,\tau\mathcal{C}^{p}\}, \;\;\;0\le p \le
N-1\\
\end{align*}
For $N$ even:
\begin{align*}
e  &  \rightarrow D_{N}\\
\mathcal{C}^{N/2}  &  \rightarrow D_{N}\\
\mathcal{C}^{p}  &  \rightarrow
{\mathchoice {\hbox{$\sf\textstyle Z\kern-0.4em Z$}} {\hbox{$\sf\textstyle Z\kern-0.4em Z$}} {\hbox{$\sf\scriptstyle
Z\kern-0.3em Z$}} {\hbox{$\sf\scriptscriptstyle Z\kern-0.2em Z$}}}%
_{N},\;\;\;1\le p \le N-1,\;p\neq\frac{N}{2}\\
\tau\mathcal{C}^{p}  &  \rightarrow\{e,\mathcal{C}^{N/2},\tau\mathcal{C}%
^{p},\tau\mathcal{C}^{p+N/2}\}, \;\;\;0\le p \le N-1\\
\end{align*}

Finally, we need to choose homomorphisms from these stabilizors into $U(1)$.
In the case of a cyclic group
${\mathchoice {\hbox{$\sf\textstyle Z\kern-0.4em Z$}}
{\hbox{$\sf\textstyle Z\kern-0.4em Z$}} {\hbox{$\sf\scriptstyle
Z\kern-0.3em Z$}} {\hbox{$\sf\scriptscriptstyle Z\kern-0.2em Z$}}}_{N}$
generated by $\mathcal{C}$, homomorphisms $\chi$ are completely determined by
$\chi(\mathcal{C})$, so that
\mbox{$\chi(\mathcal{C}^{p})=\chi(\mathcal{C})^{p}$}, with the constraint:
\mbox{$\chi(\mathcal{C})^{N}=1$}. For the group $D_{N}$ itself, we have:
\mbox{$\chi(\tau^{r}\mathcal{C}^{p})=\chi(\tau)^{r}\chi(\mathcal{C})^{p}$},
with the following constraints:
\begin{align}
\chi(\mathcal{C})^{N}  &  = 1\\
\chi(\tau)^{2}  &  = 1\\
\chi(\mathcal{C})^{2}  &  = 1
\end{align}
These are direct consequences of generator relations~(\ref{genrel1}%
),(\ref{genrel2}),(\ref{genrel3}). Again, the parity of $N$ is relevant. For
$N$ odd, \mbox{$\chi(\mathcal{C})=1$}, which leaves only two homomorphisms
from $D_{N}$ into $U(1)$. For $N$ even, \mbox{$\chi(\mathcal{C})=\pm 1$}, and
there are four such homomorphisms. The last possible stabilizor to consider is
the four element subgroup of $D_{2M}$,
\mbox{$S=\{e,\mathcal{C}^{M},\tau\mathcal{C}^{p},\tau\mathcal{C}^{p+M}\}$}.
This abelian group has four possible homomorphisms into $U(1)$, which are
characterized as follows:
\begin{align}
\chi(\mathcal{C}^{M})  &  = \pm1\\
\chi(\tau\mathcal{C}^{p})  &  = \pm1\\
\chi(\tau\mathcal{C}^{p+M})  &  = \chi(\tau\mathcal{C}^{p})\chi(\mathcal{C}%
^{M})
\end{align}

\subsection{Classification of possible models}

\label{classmodels}

\subsubsection{$N$ odd}

Let us first consider conjugacy classes of the form
\mbox{$\{\mathcal{C}^{p},\mathcal{C}^{-p}\}$}. Since the stabilizor of
$\mathcal{C}^{p}$ for the conjugacy action of $D_{N}$ is
${\mathchoice {\hbox{$\sf\textstyle Z\kern-0.4em Z$}}
{\hbox{$\sf\textstyle Z\kern-0.4em Z$}} {\hbox{$\sf\scriptstyle
Z\kern-0.3em Z$}} {\hbox{$\sf\scriptscriptstyle Z\kern-0.2em Z$}}}_{N}$, we
have: \mbox{$\chi_{\mathcal{C}^{p}}(\mathcal{C}^{q})=\omega_{p}^{q}$}, where
\mbox{$\omega_{p}^{N}=1$}. Choosing \mbox{$\chi_{\mathcal{C}^{P}}(\tau)=1$},
we have for fluxes in the cyclic group generated by $\mathcal{C}$:
\begin{align}
\chi_{\mathcal{C}^{p}}(\mathcal{C}^{q})  &  = \omega_{p}^{q}\\
\chi_{\mathcal{C}^{p}}(\tau\mathcal{C}^{q})  &  = \omega_{p}^{q}\\
\omega_{p}^{N}  &  = 1\\
\omega_{p}\omega_{-p}  &  = 1
\end{align}
For the remaining conjugacy class, we have \mbox{$\chi_{\tau}(\tau)=\eta$},
with \mbox{$\eta = \pm 1 $}. Choosing \mbox{$\chi_{\tau}(\mathcal{C}^{p})=1$},
and using Eqs.~(\ref{defchi1}) and (\ref{defchi2}), we obtain:
\begin{align}
\chi_{\tau\mathcal{C}^{p}}(\mathcal{C}^{q})  &  = 1\\
\chi_{\tau\mathcal{C}^{p}}(\tau\mathcal{C}^{q})  &  = \eta
\end{align}

All these possible phase-factors satisfy the following property:
\begin{equation}
\chi_{\Phi^{-1}}(h)\chi_{\Phi}(h)=1
\end{equation}
so that Eq.~(\ref{vacuumpaircond}) is always satisfied. So no new constraint
is imposed by the requirement to create or annihilate a pair of fluxons. What
about the stronger condition Eq.~(\ref{generalpaircond})? Its form suggests
that we should first determine pairs of fluxes \mbox{$(\Phi_{1},\Phi_{2})$}
such that their stabilizors $H_{\Phi_{1}}$ and $H_{\Phi_{1}}$ have a
non-trivial intersection. This occurs if both $\Phi_{1}$ and $\Phi_{2}$ are in
the
${\mathchoice {\hbox{$\sf\textstyle Z\kern-0.4em Z$}}{\hbox{$\sf\textstyle Z\kern-0.4em Z$}}{\hbox{$\sf\scriptstyle
Z\kern-0.3em Z$}}{\hbox{$\sf\scriptscriptstyle Z\kern-0.2em Z$}}}_{N}$ normal
subgroup generated by $\mathcal{C}$, or if
\mbox{$\Phi_{1}=\Phi_{2}=\tau\mathcal{C}^{p}$}. The second case simply implies
$\chi_{\tau}(\tau)^{2}=1$, which is not a new condition. In the first case,
choosing $h$ in
${\mathchoice {\hbox{$\sf\textstyle Z\kern-0.4em Z$}}{\hbox{$\sf\textstyle Z\kern-0.4em Z$}}{\hbox{$\sf\scriptstyle
Z\kern-0.3em Z$}}{\hbox{$\sf\scriptscriptstyle Z\kern-0.2em Z$}}}_{N}$ as well
shows that $\chi_{\Phi}(h)^{4}$ is a homomorphism from
${\mathchoice {\hbox{$\sf\textstyle Z\kern-0.4em Z$}}{\hbox{$\sf\textstyle Z\kern-0.4em Z$}}{\hbox{$\sf\scriptstyle
Z\kern-0.3em Z$}}{\hbox{$\sf\scriptscriptstyle Z\kern-0.2em Z$}}}_{N}$ to
$U(1)$ with respect to both $\Phi$ and $h$. This is satisfied in particular if
$\chi_{\Phi}(h)$ itself is a group homomorphism in both arguments. This
sufficient (but possibly not necessary) condition simplifies algebraic
considerations; it can be also justified from physical argument that the
theory should allow for the sites with different number of neighbours, $Z$,
which would change $\chi_{\Phi}(h)^{4}\rightarrow\chi_{\Phi}(h)^{Z}$.

Replacing this the constraint on $\chi_{\Phi}(h)^{4}$ by the constraint on
$\chi_{\Phi}(h)$, we get
\begin{align}
\chi_{\mathcal{C}^{p}}(\mathcal{C}^{q})  &  =\omega^{pq}\\
\omega^{N}  &  =1\\
\omega_{p}  &  =\omega^{p}%
\end{align}
Therefore, the class of possible phase-factors (which is stable under
multiplication) is isomorphic to the group \mbox{$\nbZ_{N}\times\nbZ_{2}$}.
This group of phase-factors is identical to the group of possible Chern-Simons
actions since in this case,
\mbox{$H^{3}(D_{N},U(1))=\nbZ_{N}\times\nbZ_{2}$}~\cite{Propitius95}. Very
likely, the coincidence of these two results can be traced to the absence of
projective representations for $D_{N}$ for $N$ odd~\cite{Propitius95}: as
explained above, the projective representations are not allowed in our
construction but are allowed in the classification of all possible
Chern-Simons actions \cite{Propitius95}.

\subsubsection{$N$ even}

The conjugacy classes of the form
\mbox{$\{\mathcal{C}^{p},\mathcal{C}^{-p}\}$} behave in the same way as for
$N$ odd. So writing $N=2M$, we have:
\begin{align}
\chi_{\mathcal{C}^{p}}(\mathcal{C}^{q})  &  =\omega_{p}^{q}\\
\chi_{\mathcal{C}^{p}}(\tau\mathcal{C}^{q})  &  =\omega_{p}^{q}\\
\omega_{p}^{N}  &  =1\\
\omega_{p}\omega_{-p}  &  =1\\
p  &  \notin\{0,M\},\mathrm{mod}\;2N
\end{align}
The conjugacy class $\{\mathcal{C}^{M}\}$ is special since its stabilizor is
$D_{N}$ itself. As discussed in Section \ref{genpropdihedral} above, there are
four homomorphisms form $D_{N}$ into $U(1)$ that we denote by:
\begin{align}
\chi_{\mathcal{C}^{M}}(\tau^{q}\mathcal{C}^{p})  &  =\tilde{\omega}^{q}%
\omega_{M}^{p}\\
\tilde{\omega},\omega_{M}  &  \in\{1,-1\}
\end{align}
Let us now turn to \mbox{$\chi_{\tau}(g)$} with the corresponding stabilizor
equal to \mbox{$\{e,\mathcal{C}^{M},\tau,\tau\mathcal{C}^{M}\}$}. As seen in
section \ref{genpropdihedral}, the four possible homomorphisms may be written
as:
\begin{align}
\chi_{\tau}(\tau)  &  =\eta_{0}\in\{\pm1\}\\
\chi_{\tau}(\mathcal{C}^{M})  &  =\zeta_{0}\in\{\pm1\}\\
\chi_{\tau}(\tau\mathcal{C}^{M})  &  =\eta_{0}\zeta_{0}%
\end{align}
From this, we derive the expression of $\chi_{\tau}(g)$, in the following form
(\mbox{$0 \leq p \leq M-1$}):
\begin{align}
\chi_{\tau}(\mathcal{C}^{p})  &  =1\\
\chi_{\tau}(\mathcal{C}^{p+M})  &  =\zeta_{0}\\
\chi_{\tau}(\tau\mathcal{C}^{-p})  &  =\eta_{0}\\
\chi_{\tau}(\tau\mathcal{C}^{-p+M})  &  =\eta_{0}\zeta_{0}%
\end{align}
Furthermore, because
\mbox{$\mathcal{C}^{p}\tau\mathcal{C}^{-p}=\tau\mathcal{C}^{-2p}$},
Eq.~(\ref{defchi2}) implies:
\begin{equation}
\chi_{\tau\mathcal{C}^{-2p}}(g)=\chi_{\tau}(g\mathcal{C}^{p})
\end{equation}
The last conjugacy class to consider contains $\tau\mathcal{C}$, with the
stabilizor
\mbox{$\{e,\mathcal{C}^{M},\tau\mathcal{C},\tau\mathcal{C}^{1+M}\}$}. In this
case, we may set (\mbox{$0 \leq p \leq M-1$}):
\begin{align}
\chi_{\tau\mathcal{C}}(\mathcal{C}^{p})  &  =1\\
\chi_{\tau\mathcal{C}}(\mathcal{C}^{p+M})  &  =\zeta_{1}\in\{\pm1\}\\
\chi_{\tau\mathcal{C}}(\tau\mathcal{C}^{1-p})  &  =\eta_{1}\in\{\pm1\}\\
\chi_{\tau\mathcal{C}}(\tau\mathcal{C}^{1-p+M})  &  =\eta_{1}\zeta_{1}\\
\chi_{\tau\mathcal{C}^{1-2p}}(g)  &  =\chi_{\tau\mathcal{C}}(g\mathcal{C}^{p})
\end{align}

Here again, the constraint Eq.~(\ref{vacuumpaircond}) is always satisfied. To
impose Eq.~(\ref{generalpaircond}), we have to consider pairs of fluxes
\mbox{$(\Phi_{1},\Phi_{2})$} such that $H_{\Phi_{1}}\cap H_{\Phi_{2}}$ is
non-trivial. As before, choosing $\Phi_{1}$ and $\Phi_{2}$ in the
${\mathchoice {\hbox{$\sf\textstyle Z\kern-0.4em Z$}}
{\hbox{$\sf\textstyle Z\kern-0.4em Z$}} {\hbox{$\sf\scriptstyle
Z\kern-0.3em Z$}} {\hbox{$\sf\scriptscriptstyle Z\kern-0.2em Z$}}}_{N}$
subgroup imposes $\omega_{p}=\omega^{p}$, with $\omega^{N}=1$. A new
constraint arises by choosing $\Phi_{1}=\mathcal{C}^{p}$ and $\Phi_{2}%
=\tau\mathcal{C}^{p^{\prime}}$. In this case, $\mathcal{C}^{M}$ belongs to
their common stabilizor. Eq.~(\ref{generalpaircond}) implies:
\begin{equation}
\chi_{\mathcal{C}^{p}}(\mathcal{C}^{M})\chi_{\tau\mathcal{C}^{p^{\prime}}%
}(\mathcal{C}^{M}) =\chi_{\mathcal{C}^{p}\tau\mathcal{C}^{p^{\prime}}%
}(\mathcal{C}^{M})=\chi_{\tau\mathcal{C}^{p^{\prime}+p}}(\mathcal{C}^{M})
\label{Ctauconstraint}%
\end{equation}
But using Eq.~(\ref{defgaugegroup}) this yields:
\begin{equation}
\chi_{\tau\mathcal{C}^{p+p^{\prime}}}(\mathcal{C}^{M+p})=\chi_{\tau
\mathcal{C}^{p+p^{\prime}}}(\mathcal{C}^{M}) \chi_{\tau\mathcal{C}%
^{p+p^{\prime}}}(\mathcal{C}^{p})
\end{equation}
Therefore we have the constraint: \mbox{$\zeta_{0}=\zeta_{1}=1$}, which
enables us to simplify drastically the above expression for phase-factors:
\begin{align}
\chi_{\tau\mathcal{C}^{2p}}(\mathcal{C}^{q})  &  = 1\\
\chi_{\tau\mathcal{C}^{2p}}(\tau\mathcal{C}^{q})  &  = \eta_{0}%
\end{align}
and
\begin{align}
\chi_{\tau\mathcal{C}^{2p+1}}(\mathcal{C}^{q})  &  = 1\\
\chi_{\tau\mathcal{C}^{2p+1}}(\tau\mathcal{C}^{q})  &  = \eta_{1}%
\end{align}
But Eq.~(\ref{Ctauconstraint}) now implies that $\chi_{\mathcal{C}^{p}%
}(\mathcal{C}^{M})=1$ for any $p$, which is satisfied only when $\omega^{M}%
=1$. Specializing to $p=M$, we see that the common stabilizor contains now two
more elements, namely $\tau\mathcal{C}^{p^{\prime}}$ and $\tau\mathcal{C}%
^{p^{\prime}+M}$. Eq.~(\ref{Ctauconstraint}) now requires that:
\begin{align}
M\;\mathrm{even}  &  \rightarrow\tilde{\omega}=1\\
M\;\mathrm{odd}  &  \rightarrow\tilde{\omega}\eta_{0}\eta_{1}=1
\end{align}
It is then easy to check that considering the common stabilizor of
$\tau\mathcal{C}^{p}$ and $\tau\mathcal{C}^{p^{\prime}}$, which may contain
two or four elements, does not bring any new constraint. Finally, $\omega$
belongs to the ${\mathchoice {\hbox{$\sf\textstyle Z\kern-0.4em Z$}}
{\hbox{$\sf\textstyle Z\kern-0.4em Z$}} {\hbox{$\sf\scriptstyle
Z\kern-0.3em Z$}} {\hbox{$\sf\scriptscriptstyle Z\kern-0.2em Z$}}}_{M}$ group
and among the three binary variables $\eta_{0}$, $\eta_{1}$, and
$\tilde{\omega}$, only two are independent. Therefore, the set all all
possible phase-factors for the $D_{2M}$ group is identical to
\mbox{$\nbZ_{M} \times \nbZ_{2} \times \nbZ_{2}$}. This contains only half of
$H^{3}(D_{2M},U(1))$ which is equal to
\mbox{$\nbZ_{2M} \times \nbZ_{2} \times \nbZ_{2}$}~\cite{Propitius95}. But
$D_{2M}$ admits non trivial projective representations since $H^{2}%
(D_{2M},U(1))={\mathchoice {\hbox{$\sf\textstyle Z\kern-0.4em Z$}}
{\hbox{$\sf\textstyle Z\kern-0.4em Z$}} {\hbox{$\sf\scriptstyle
Z\kern-0.3em Z$}} {\hbox{$\sf\scriptscriptstyle Z\kern-0.2em Z$}}}_{2}$ which
cannot appear in our construction. The important result is that in spite of
this restriction, we get a non-trivial subset of the possible theories also in
this case.

\subsection{Holonomy properties}

Using the general expression~(\ref{adiabaticphase}), and the above description
of possible phase-factors, we may compute the adiabatic phase induced by a
process where a fluxon $g_{1}$ winds once around another fluxon $g_{2}$. The
results are listed in the last column of table~\ref{finaltable}, and they are
valid for \emph{both} parities of $N$.

\begin{table}[th]
\caption{Adiabatic phase $f(g_{1},g_{2})$ generated in the process where
fluxon $g_{1}$ winds once around a fluxon $g_{2}$. Values of phase-factors
$\chi_{g^{\prime}_{1}}(g^{\prime}_{1})$ and $\chi_{g^{\prime}_{2}}(g^{\prime
}_{1})$ are given for dihedral groups $D_{N}$ with odd $N$. For even $N$,
expressions for $\chi_{g^{\prime}_{1}}(g^{\prime}_{1})$ and $\chi_{g^{\prime
}_{2}}(g^{\prime}_{1})$ are slightly more complicated, but interestingly, the
main result for $f(g_{1},g_{2})$ is the same as for odd $N$, namely it
involves only the complex number $\omega$.}%
\label{finaltable}%
\begin{tabular}
[c]{|c|c|c|c||c|c|c|}\hline
$g_{1}$ & $g_{2}$ & $g^{\prime}_{1}$ & $g^{\prime}_{2}$ & $\chi_{g^{\prime
}_{1}}(g^{\prime}_{1})$ & $\chi_{g^{\prime}_{2}}(g^{\prime}_{1})$ &
$f(g_{1},g_{2})$\\\hline\hline
$\mathcal{C}^{p}$ & $\mathcal{C}^{q}$ & $\mathcal{C}^{p}$ & $\mathcal{C}^{q}$
& $\omega^{p^{2}}$ & $\omega^{pq}$ & $\omega^{2p(p+2q)}$\\\hline
$\tau\mathcal{C}^{p}$ & $\mathcal{C}^{q}$ & $\tau\mathcal{C}^{p+2q}$ &
$\mathcal{C}^{-q}$ & $\eta$ & $\omega^{-q(p+2q)}$ & $\omega^{-4q(p+2q)}%
$\\\hline
$\mathcal{C}^{p}$ & $\tau\mathcal{C}^{q}$ & $\mathcal{C}^{-p}$ &
$\tau\mathcal{C}^{q-2p}$ & $\omega^{p^{2}}$ & $1$ & $\omega^{2p^{2}}$\\\hline
$\tau\mathcal{C}^{p}$ & $\tau\mathcal{C}^{q}$ & $\tau\mathcal{C}^{-p+2q}$ &
$\tau\mathcal{C}^{3q-2p}$ & $\eta$ & $\eta$ & $1$\\\hline
\end{tabular}
\end{table}

\section{Conclusion}

\bigskip Generally, in order to appear as a low energy sector of some physical
Hamiltonian, the Chern-Simons gauge theory has to involve gauge
transformations that depend only on a local flux configuration. Furthermore,
to be interesting from the view point of quantum computation, the theory
should allow for a local gauge invariant electric field operator that moves a
flux or fuses two fluxes together. Here we have analyzed non-Abelian gauge
models that satisfy these general conditions; our main result is the equation
(\ref{generalpaircond}) for the phase factor $\chi$ associated with a gauge
transformation. Furthermore, we have computed the flux braiding properties for
a given phase factor that satisfies these conditions. Finally, we have applied
our general results to the simplest class of non-Abelian groups, dihedral
groups $D_{n}$. The fluxon braiding properties in these groups are summarized
in Table I.

Inspection of the Table I shows that even for the smallest groups,
Chern-Simons term modifies braiding properties in a very non-trivial manner
and gives phases to the braiding that were trivial in the absence of the
Chern-Simons term. In the scheme\cite{Mochon2004} where the pair of two
compensating fluxes $(\tau\mathcal{C}^{p},\tau\mathcal{C}^{p})$ are used to
encode one bit, the transformations allowed in the absence of Chern-Simons
term are limited to conjugation. In the presence of Chern-Simons term the
braiding of such bit with the controlled flux results in a richer set of
transformations that involve both conjugation by group elements and phase
factors (see Table I) but does not change the state of the flux as it should.
We hope that this will make it possible to construct the universal quantum
computation with the simplest group $D_{3}$ that does not involve operations
that are difficult to protect (such as charge doublets).

The implementation of the microscopic Hamiltonians discussed in this paper in
a realistic physical system is a challenging problem. The straightforward
implementation would mean that the dominant term in the microscopic
Hamiltonian is $H=-t\sum_{i,g}U_{i}(g)$ so that all low energy states are
gauge invariant. This is not easy to realize in a physical system because
operator $U(g)$ involves a significant number of surrounding bonds. We hope,
however, this can be achieved by a mapping to the appropriate spin model as
was the case for Abelian Chern-Simons theories; this is the subject of the
future research.

\textbf{Acknowledgments}

We are thankful to M. G\"orbig and R. Moessner for useful discussions. LI is
thankful to LPTHE, Jussieu for their hospitality while BD has enjoyed the
hospitality of the Physics Department at Rutgers University. This work was
made possible by support from NATO CLG grant 979979, NSF DMR 0210575.

\appendix

\section{Electric field gauge invariance: general criterion}

\label{amplsol}

\begin{figure}[th]
\includegraphics[width=3.0in]{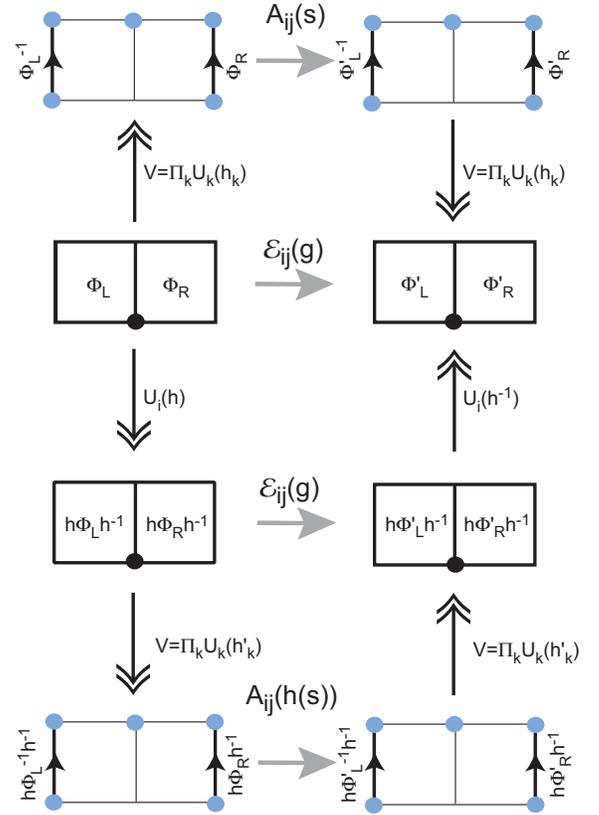}\caption{(Color
online) Derivation of the general condition on the phase factors from the
requirement of electric gauge field invariance. The action of the electric
field on arbitrary states (middle of the figure) related to each other by the
gauge transformation at site $i$ (shown as black dot) can be expressed via the
phase factors $A(s)$ and $A(h(s))$ characterizing the transition between
reference states (bottom and top). This involves gauge transformations on all
grey (blue) sites. The gauge invariance implies that the total phase factor
accumulated when moving around the diagram is unity which implies the
condition on the ratio $A(s)/A(h(s))$. As in text, $s$ stands for a triple
$(\Phi_{L},\Phi_{R},g)$, and $h(s)$ is obtained by conjugating the components
of $s$ by $h$. }%
\label{ElectricFieldGaugeInvarianceGeneral}%
\end{figure}

In the absence of additional constraints on the gauge transformation phase
factors, $\chi_{\Phi}(g)$, the definition of the electric field operator
implies that it is gauge invariant under gauge transformations on all sites
except $i$ while the the gauge transformation on the latter gives
\begin{equation}
U_{i}(h)\mathcal{E}_{ij}(g)|\Psi\rangle=\alpha(\Phi_{L},\Phi_{R}%
,g|h)\mathcal{E}_{ij}(hgh^{-1})U_{i}(h)|\Psi\rangle
\end{equation}
with
\begin{align}
\alpha(\Phi_{L},\Phi_{R},g|h)  &  =\left(  \frac{\chi_{g\Phi_{L}}(h)}%
{\chi_{\Phi_{L}}(h)}\frac{\chi_{\Phi_{R}g^{-1}}(h)}{\chi_{\Phi_{R}}%
(h)}\right)  ^{4}\\
&  \times\frac{A(\Phi_{L},\Phi_{R},g)}{A(h\Phi_{L}h^{-1},h\Phi_{R}%
h^{-1},hgh^{-1})}%
\end{align}
Here the notations are those of Fig.~\ref{ElectricFieldConstruction}, where
state $|\Psi\rangle$ corresponds to the initial state on the figure. The
computation leading to this equation is depicted on
Fig~\ref{ElectricFieldGaugeInvarianceGeneral}. The question we have to address
now is whether it is possible to choose amplitudes $A(\Phi_{L},\Phi_{R},g)$ so
that the function $\alpha(\Phi_{L},\Phi_{R},g|h)$ is equal to unity.

To simplify notations, let us denote triples \mbox{$(\Phi_{L},\Phi_{R},g)$} by
a single label $s$. The group $G$ acts on these triples according to:
\begin{equation}
s\equiv(\Phi_{L},\Phi_{R},g)\overset{h}{\longrightarrow}(h\Phi_{L}h^{-1}%
,h\Phi_{R}h^{-1},hgh^{-1})\equiv h(s)
\end{equation}
Let us also introduce the function $f(s|h)$ defined as:
\begin{equation}
f(s|h)=\left(  \frac{\chi_{g\Phi_{L}}(h)}{\chi_{\Phi_{L}}(h)}\frac{\chi
_{\Phi_{R}g^{-1}}(h)}{\chi_{\Phi_{R}}(h)}\right)  ^{4}.
\end{equation}
The condition $\alpha(s|h)=1$ now reads:
\begin{equation}
\frac{A(h(s))}{A(s)}=f(s|h) \label{ampleq}%
\end{equation}
Now we show that solutions $A(s)$ exist if and only if the function $f(s|h)$
satisfies the following conditions:
\begin{align}
f(s|h)  &  =1\;\;\mathrm{if}\;\;h(s)=s\label{condstab}\\
f(s|h_{1}h_{2})  &  =f(h_{2}(s)|h_{1})f(s|h_{2}) \label{grouprel}%
\end{align}
Note that the second condition is very similar to the fundamental
property~(\ref{defgaugegroup}) which enforces the existence of a local gauge
symmetry group. In fact, using~(\ref{defgaugegroup}), we can check that
property~(\ref{grouprel}) always holds. Property~(\ref{condstab}) is a new and
quite strong condition imposed on gauge phase-factors $\chi_{\Phi}(g)$; it is
equivalent to the condition (\ref{PhaseFactorCondition}) which in the main
text has been derived as \emph{necessary}. We now show that this condition is
also \emph{sufficient}.

Let us consider the equation:
\begin{equation}
\label{ampleqap}\frac{A(g(s))}{A(s)}=f(s|g)
\end{equation}
where $f(s|g)$ and $A(s)$ are complex numbers with unit modulus, $s$ belongs
to a set $S$ on which the group $G$ acts, and $g\in G$. Here, the function
$f(s|g)$ is supposed to be known, and we are looking for ``amplitudes''
$A(s)$. Let us denote by $H_{s}$ the stabilizor of $s$, namely it is composed
of all the group elements $h$ such that \mbox{$h(s)=s$.} From
Eq.~(\ref{ampleqap}), we get the two conditions:
\begin{align}
f(s|h)  &  = 1\;\;\mathrm{if}\;\;h(s)=s\label{conda}\\
f(s|g_{1}g_{2})  &  = f(g_{2}(s)|g_{1})f(s|g_{2}) \label{condb}%
\end{align}
The second equation is obtained from:
\begin{equation}
\frac{A(g_{1}g_{2}(s))}{A(s)}=\frac{A(g_{1}g_{2}(s))}{A(g_{2}(s))}
\frac{A(g_{2}(s))}{A(s)}.
\end{equation}
Note that these two conditions imply that:
\begin{equation}
f(s|gh)=f(s|g)\;\;\;\mathrm{when}\;\;\;h\in H_{s} . \label{condc}%
\end{equation}

Let us now show that when these two conditions are satisfied, we can always
reconstruct a system of amplitudes solving Eq.~(\ref{ampleqap}). From the form
of this equation, we see that the various orbits in $S$ for the action of $G$
remain uncoupled. Let us then concentrate on one orbit generated by a fixed
given element $s$. Elements of this orbit are in one to one correspondence
with the left cosets $gH_{s}$ of $G$, since $gh(s)=g(s)$ when $h\in H_{s}$.
Let us choose in each coset a representative element $g_{n}$. Our problem is
to find amplitudes $A(g_{n}(s))$ knowing $A(s)$. Let us define:
\begin{equation}
A(g_{n}(s))\equiv A(s)f(s|g_{n})
\end{equation}
From Eq.~(\ref{condc}), we have \mbox{$A(g(s))=A(s)f(s|g)$} for any $g$ in
$G$, since $g$ may always be written as $g_{n}h$ with $h\in H_{s}$. Now
Eq.~(\ref{ampleqap}) is an immediate consequence of the condition~(\ref{condb}).

To conclude, when conditions~(\ref{conda}) and~(\ref{condb}) are satisfied,
solutions of Eq.~(\ref{ampleqap}) can be constructed independently on each
orbit of $S$ for the action of $G$. On a given orbit, the solution is unique,
up to the choice of $A(s)$ for one particular element of this orbit.

\section{Structure of the ground state on a torus}

\label{sectorus}

The distinguishing feature of a torus is the appearance of non-trivial closed
loops that are classified according to their winding numbers associated with
two fundamental cycles $\gamma_{x}$ and $\gamma_{y}$, chosen with a common
origin $O$. As a consequence, even if local fluxes vanish on all plaquettes,
the fluxes $\Phi_{x}$ and $\Phi_{y}$ associated to $\gamma_{x}$ and
$\gamma_{y}$ may not vanish. These global degrees of freedom are the origin of
the topological degeneracies exhibited by a large class of gauge-invariant
models on a closed space with non-trivial topology~\cite{Wen90}. How are these
degeneracies affected by the presence of a Chern-Simons term? Precise formulas
for the Hilbert-space dimension of the \emph{pure} Chern-Simons theory with a
finite gauge group $G$ have been given~\cite{Dijkgraaf1990} in terms of the
action via functions \mbox{$\alpha(h,k,l)$} in $H^{3}(G,U(1))$. In this paper
we consider a much larger Hilbert space induced by all allowed classical gauge
configurations, $|\{g_{ij}\}\rangle$. We may expect to recover the pure
Chern-Simons Hilbert-space (which dimension is independent of the system size)
by projecting the full Hilbert-space onto the gauge-invariant ground-state of
some local gauge-invariant Hamiltonian. This goes beyond the scope of the
present paper because our goal here is classification and basic properties of
such local gauge invariant Hamiltonian. However, it is instructive to
understand how, in the present formalism, a Chern-Simons term affects the
gauge dynamics in the topological fluxless sector on a torus.

\begin{figure}[th]
\includegraphics[width=2.0in]{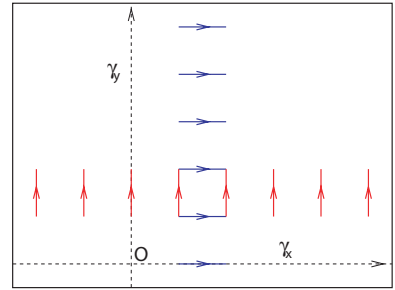} \caption{(Color online) For a
square lattice on a torus, fundamental oriented cycles $\gamma_{x}$ and
$\gamma_{y}$ are depicted as black dashed lines. The vertical string of
horizontal blue lines represents links carrying the group element $\Phi_{x}$.
Similarly, the horizontal string of vertical red lines represents links
carrying $\Phi_{y}$. On all other links $ij$ on the square lattice $g_{ij}=e$,
so we have not represented them not to overcrowd the figure. }%
\label{figtorustemplate}%
\end{figure}

Starting from the trivial configuration $g_{ij}=e$, we may induce a new
fluxless state in the bulk but with a non-trivial $\Phi_{x}$ along $\gamma
_{x}$ by creating a string of parallel horizontal links carrying each the
group element $\Phi_{x}$ as shown on Fig.~\ref{figtorustemplate}. This string
may be viewed as the result of a succession of elementary processes discussed
in section~\ref{sectionprocesses}: first the creation of a $\Phi_{x}$,
$\Phi_{x}^{-1}$ fluxon pair, then the motion of one of them along $\gamma_{y}$
and finally annihilation of the two fluxons. As we have seen, each of these
processes may be described by a gauge-invariant operator. Let us denote by
$\mathcal{C}_{x}(\Phi_{x})$ the product of all these operators involved in the
creation of the string. Similarly, we define $\mathcal{C}_{y}(\Phi_{y})$. More
generally, we may start from a state already characterized by a pair of fluxes
$(\Phi_{x},\Phi_{y})$. Since $\gamma_{x}$ and $\gamma_{y}$ commute up to
homotopy, we require that $\Phi_{x}$ and $\Phi_{y}$ commute in $G$. The string
picture allows us to construct operators $\mathcal{C}_{x}(g_{x})$ and
$\mathcal{C}_{y}(g_{y})$ such that:
\begin{align}
(\Phi_{x},\Phi_{y})  &  \overset{\mathcal{C}_{x}(g_{x})}{\longrightarrow
}(g_{x}\Phi_{x},\Phi_{y})\\
(\Phi_{x},\Phi_{y})  &  \overset{\mathcal{C}_{y}(g_{y})}{\longrightarrow}%
(\Phi_{x},g_{y}\Phi_{y})
\end{align}
Denoting by $|\Phi_{x},\Phi_{y}\rangle$ any state with the pair of fluxes
$(\Phi_{x},\Phi_{y})$, we show below that $\mathcal{C}_{x}(g_{x})$ and
$\mathcal{C}_{y}(g_{y})$ do not commute, but their actions on a state
$|\Phi_{x},\Phi_{y}\rangle$ are related by phase-factors:
\begin{align}
\mathcal{C}_{y}(g_{y})\mathcal{C}_{x}(g_{x})|\Phi_{x},\Phi_{y}\rangle &
=\lambda(\Phi_{x},\Phi_{y}|g_{x},g_{y})\nonumber\\
&  \times\mathcal{C}_{x}(g_{x})\mathcal{C}_{y}(g_{y})|\Phi_{x},\Phi_{y}%
\rangle\label{defininglambda}\\
\lambda(\Phi_{x},\Phi_{y}|g_{x},g_{y})  &  =\frac{\chi_{g_{x}}(g_{y})^{2}%
\chi_{\Phi_{x}^{-1}g_{x}\Phi_{x}}(g_{y})^{2}}{\chi_{g_{y}^{-1}}(g_{x})^{2}%
\chi_{\Phi_{y}^{-1}g_{y}^{-1}\Phi_{y}}(g_{x})^{2}} \label{eqphasestrings}%
\end{align}
Note that in order for all four states \mbox{$|\Phi_{x},\Phi_{y}\rangle$},
\mbox{$|g_{x}\Phi_{x},\Phi_{y}\rangle$},
\mbox{$|\Phi_{x},g_{y}\Phi_{y}\rangle$},
\mbox{$|g_{x}\Phi_{x},g_{y}\Phi_{y}\rangle$} to be defined, the following
constraints have to be imposed:
\begin{align}
\Phi_{x}\Phi_{y}  &  =\Phi_{y}\Phi_{x}\\
\Phi_{x}g_{y}  &  =g_{y}\Phi_{x}\\
g_{x}\Phi_{y}  &  =\Phi_{y}g_{x}\\
g_{x}g_{y}  &  =g_{y}g_{x}%
\end{align}
This relation~(\ref{eqphasestrings}) is reminiscent of the phase factors
involved in permuting electrical field operators attached to adjacent links
for Abelian models with a Chern-Simons term~\cite{Doucot2005,Doucot2005b}.

\begin{widetext}
\begin{figure*}[ht]
\includegraphics[width=5.0in]{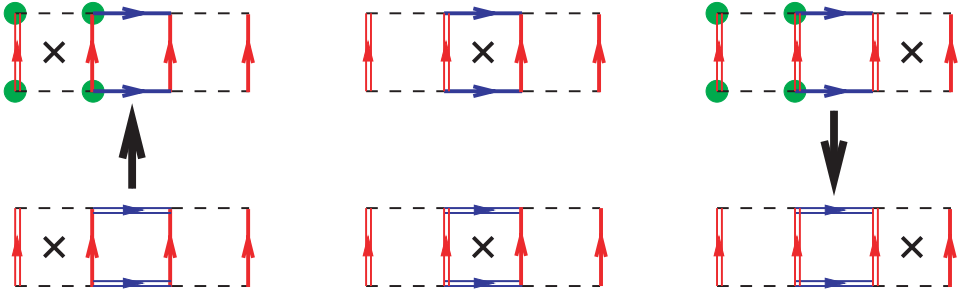}\caption{(Color online) The
upper line illustrates intermediate steps in the action of $\mathcal{C}
_{y}(g_{y})$ on the template $|\Phi_{x},\Phi_{y}\rangle$. The plain
(double) vertical red lines carry the element $\Phi_{y}$ ($g_{y}\Phi_{y}
$) whereas the plain horizontal blue lines carry $\Phi_{x}$. Successive
locations of the fluxon moving from left to right are represented by a black
cross. The lower line describes the action of the same string operator on the
new state $\mathcal{C}_{x}(g_{x})|\Phi_{x},\Phi_{y}\rangle$ and the double
horizontal blue lines carry now $g_{x}\Phi_{x}$. One goes from the upper to
the lower line by a gauge transformation attached to the sites indicated by a
green circle. Since this transformation commutes with the fluxon motion
operators, these processes generate a contribution to
$\lambda(\Phi_{x},\Phi_{y}|g_{x},g_{y})$ defined in section~\ref{sectorus} equal to the product
of the phase factors due to the two gauge transformations shown by thick black
arrows. This factor is equal to
$\left(  \chi_{g_{y}^{-1}}(g_{x})\chi_{\Phi_{y}^{-1}g_{y}^{-1}\Phi_{y}}(g_{x})\right)^{-2}$.
A second contribution to $\lambda(\Phi_{x},\Phi_{y}|g_{x},g_{y})$ comes from the
comparison between the actions of $\mathcal{C}_{x}(g_{x})$ on
$|\Phi_{x},\Phi_{y}\rangle$ and \mbox{$\mathcal{C}_{y}(g_{y})|\Phi_{x},\Phi_{y}\rangle$}.
This process is not shown on the figure, and yields a term equal to
$\left(  \chi_{g_{x}}(g_{y})\chi_{\Phi_{x}^{-1}g_{x}\Phi_{x}}(g_{y})\right)^{2}$}
\label{figstringcreation}
\end{figure*}
\end{widetext}

Let us now briefly explain how to derive the relations~(\ref{defininglambda})
and (\ref{eqphasestrings}) expressing the commutation rules between
$\mathcal{C}_{x}(g_{x})$ and $\mathcal{C}_{y}(g_{y})$. The first step is to
choose template states for a pair of fluxes $(\Phi_{x},\Phi_{y})$ along
elementary cycles $\gamma_{x}$ and $\gamma_{y}$. These are depicted on
Fig.~\ref{figtorustemplate}. Let us now compare the action of
\mbox{$\mathcal{C}_{x}(g_{x})\mathcal{C}_{y}(g_{y})$} and
\mbox{$\mathcal{C}_{y}(g_{y})\mathcal{C}_{x}(g_{x})$} on this template
\mbox{$|\Phi_{x},\Phi_{y}\rangle$}. The situation is illustrated on
Fig.~\ref{figstringcreation}. Note that we do not need explicitely the values
of the fluxon hopping amplitudes $A(\Phi_{L},\Phi_{R},g)$, but simply the fact
they are local and that fluxon moving operators commute with gauge
transformation. The value of $\lambda(\Phi_{x},\Phi_{y}|g_{x},g_{y})$ obtained
in this way (see Eq.~(\ref{eqphasestrings})) is valid not only for our
template state $|\Phi_{x},\Phi_{y}\rangle$ but for any state deduced from it
by a gauge transformation.

\end{document}